\documentclass[aps,prb,superscriptaddress,twocolumn,reprint,floatfix,
]{revtex4-2}
\usepackage{amsfonts}
\usepackage{amstext}
\usepackage{amssymb}
\usepackage{amsmath}
\usepackage[utf8]{inputenc}
\usepackage{graphicx}
\usepackage{color,xcolor}
\usepackage{hyperref}
\hypersetup{colorlinks=true,%
            citecolor=blue,
            linkcolor=blue,
            filecolor=blue,
            urlcolor=blue,
            pdfstartview=FitH,
            pdfpagemode=UseNone}
\usepackage{tikz}
\graphicspath{ {figures/} }

\begin{document}

\title{Local equilibrium charge and spin currents in two-dimensional
topological  systems}

\author{Leandro R. F. Lima}
\affiliation{Departamento de F\'{\i}sica, Instituto de Ci\^encias Exatas, Universidade Federal Rural do Rio de Janeiro, 23897-000 Serop\'edica - RJ, Brazil}

\author{Caio Lewenkopf}
\affiliation{Instituto de F\'{\i}sica, Universidade Federal Fluminense, 24210-346 Niter\'oi - RJ, Brazil}

\date{\today}

\begin{abstract}
We study the equilibrium and nonequilibrium electronic transport properties of multiprobe topological systems using a combination of the Landauer-B\"uttiker approach and nonequilibrium Green's functions techniques. We obtain general expressions for both nonequilibrium and equilibrium local electronic currents that, by suitable projections, allow one to compute charge, spin, valley, and orbital currents. We show that external magnetic fields give rise to equilibrium charge currents in mesoscopic system and study the latter in the quantum Hall regime. Likewise, a spin-orbit interaction leads to local equilibrium spin currents, that we analyze in the quantum spin Hall regime. We show that an accurate theoretical assessment of the equilibrium currents is quite challenging and propose a two-measurement protocol that facilitates a comparison between experiment and theory. 
\end{abstract}

\maketitle

\section{Introduction}
\label{sec:intro}

Edge states and the bulk-boundary correspondence are key features of systems with topological properties \cite{Hasan2010,Qi2011,Bernevig2013}. Large experimental and theoretical interest has been devoted 
to electronic two-dimensional (2D) topological systems, which are characterized by edge states that are robust against disorder and by a quantized conductance. 
Currently, it is well established that  the current flow in integer quantum Hall (IQH) 
and quantum spin Hall (QSH) systems  is associated with gapless chiral edge states \cite{Halperin1982} and helical edge 
states \cite{Kane2005}, respectively. 

Advances in device fabrication and detection techniques allow one to measure edge currents in IQH \cite{vanHaren1995,McCormick1999,Yacoby1999,Suddards2012,Pasher2014} and QSH systems \cite{Roth2009,Nowack2013,Spanton2014}, using a variety of methods. As a result, it is nowadays possible to experimentally assess local current maps of 2D samples \cite{Nowack2013,Chang2017,Tetienne2017,Shi2019}, whose interpretation calls for a state-of-the-art microscopic electronic transport theory.  

Of particular interest is the investigation of dissipationless currents proposed in both in IQH \cite{Geller1995} and QSH systems \cite{Buttiker2009, Sonin2011, Ando2013, Maekawa2017}. 
While recent current flow quantum imaging experiments in graphene \cite{Tetienne2017,Uri2020} have rekindled the interest in equilibrium currents in IQH systems \cite{Lee2004}, to the best of our knowledge experimental evidence of equilibrium currents in QSH systems is still missing. 

Equilibrium electronic currents in (nontopological) mesoscopic conductors with broken time-reversal symmetry have been theoretically addressed a long time ago \cite{Baranger1989} and gained a substantial attention \cite{Buttiker1983,Altshuler1991,MullerGroeling1993} due to experiments on persistent currents in mesoscopic rings \cite{Levy1990,Bluhm2009,Bleszynski-Jayich2009}.

However, a microscopic study of equilibrium currents in the {IQH} regime in a multiprobe set-up is still lacking. 
The situation is less clear for QSH systems.
Equilibrium currents in topological insulators have been studied for massless \cite{Mishchenko2014} as well as for massive Dirac electrons \cite{Silvestrov2019,Chen2020}. Interestingly, these works overlook the literature on spin currents in the spin Hall regime. 
The existence of bulk spin currents in thermodynamic equilibrium in conductors lacking inversion symmetry have been first discussed more than 15 years ago by Rashba \cite{Rashba2003} and few years later in more general terms \cite{Tokatly2008}. Application proposals of equilibrium spin currents \cite{Pareek2004} have been subject of controversy, until it was shown that both the Landauer-B\"uttiker approach \cite{Kiselev2005} and NEGF theory \cite{Souma2005,Nikolic2006} are incompatible with a net equilibrium spin transport.

We address the problem of equilibrium currents using the nonequilibrium Green's functions theory (NEGF) \cite{Jauho2008} combined with the Landauer-B\"uttiker approach \cite{Landauer1970, Buttiker1985} that allows one to obtain the local transport properties of a given multiprobe system with arbitrary geometry. We employ the formalism put forward in Refs.~\cite{Cresti2003,Nikolic2006} to compute local electronic and spin currents using the recursive Green's 
functions (RGF) method  \cite{MacKinnon1985,Lewenkopf2013,Lima2016,Lima2018}. 
We use this approach  to study the charge and the spin flow in topological systems. 
We show that the equilibrium currents cannot be measured by the standard methods and discuss strategies to assess them. In addition, our analysis also shows the necessity of separating the equilibrium from the nonequilibrium contributions to the local currents for a correct theoretical interpretation of  experimental results. While nonequilibrium transport is governed by the properties of the Fermi surface states, equilibrium currents involve all occupied states. Hence, their quantitative assessment is theoretically quite daunting, since it requires not only accounting for electronic states deep into the Fermi sea, but also for precise description of the system geometry, disorder configuration, and material band structure. To circumvent this issue, we put forward a simple two-measurement protocol that facilitates a quantitative comparison between theory and experiments.

In summary, we show that the equilibrium currents in topological systems, contrary to the common believe, do not stem solely from the (topological) edge states. They actually have a significant contribution from electronic states that belong to the trivial phase, which is quite challenging to calculate with a 
good level of accuracy for realistic models. We propose a simple measurement protocol to separate both contributions, allowing for an amenable computation, as well as for an unambiguous experimental assessment of the (topological) edge state equilibrium currents.

This paper is structured as follows. In Sec.~\ref{sec:multiprobe}, we define the problem and obtain close expressions for the local charge and spin currents in a multiprobe set-up using nonequilibrium Green's functions. We identify the equilibrium and nonequilibrium current components and discuss their properties in terms of the Landauer-B\"uttiker approach. In Sec.~\ref{sec:toymodel} we introduce an analytical solvable model to show the necessity of an external magnetic field to generate equilibrium charge currents. In Sec.~\ref{sec:IQH} we study the nonequilibrium and the equilibrium currents in the integer quantum Hall regime for a realistic multiterminal setup. In Sec.~\ref{sec:QSH} we conduct a similar analysis for quantum spin Hall systems. We summarize our results and present our conclusions in Sec.~\ref{sec:conclusions}.

\section{Electronic current in a multiprobe mesoscopic system}
\label{sec:multiprobe}

We begin this section with a brief overview of the Landauer-B\"uttiker approach \cite{Landauer1970, Buttiker1985} 
for the description of the transport properties of multiprobe quantum coherent mesoscopic systems.
These results, originally obtained using the scattering matrix approach \cite{Buttiker1986,Blanter2000}, can be derived using nonequilibrium Green's functions (NEGF) \cite{Datta1995,Hernandez2013}. In this paper we adopt the latter.
Next, we discuss the theory that describes local equilibrium and nonequilibrium electronic currents \cite{Todorov2002, Cresti2003, Nikolic2006, Zarbo2007}. We cast the corresponding expressions in a convenient way to implement an efficient recursive Green's function (RGF) method \cite{Lewenkopf2013,Lima2016,Lima2018} that is used to compute the results reported in Secs.~\ref{sec:IQH} and \ref{sec:QSH}.

We consider the standard model Hamiltonian
\begin{align}
	H=H_C+H_L+V,
	\label{hamiltonian}
\end{align}
where $H_C$ describes the mesoscopic system $C$, $H_L$ the leads that connect $C$ to electronic reservoirs in thermal and chemical equilibrium, and $V$ the coupling between the mesoscopic system $C$ and the leads. 

\begin{figure}[t]
\vskip0.2cm
	\centering
		\includegraphics[width=0.75\columnwidth]{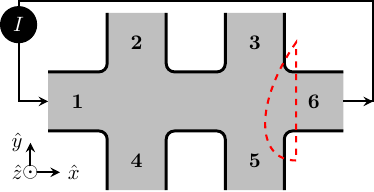}
\caption{
Sketch of a multiterminal Hall bar. 
The terminals (or leads) $L$, labeled by $\alpha=1,\cdots, 6$, can inject (collect) electrons into (from) the system $C$.
The dashed line illustrates a possible system control cross section and $I$ indicates the current path used in the numerical calculations (see text).
}
	\label{fig:hall_bars_setup_A}
\end{figure}

For the sake of concreteness, we consider a Hall bar in Fig.~\ref{fig:hall_bars_setup_A}.
The multiterminal Landauer-B\"uttiker formula for the electronic current $I_\alpha$ at the terminal
$\alpha$ reads \cite{Buttiker1986,Datta1995}
\begin{align}
	I_\alpha
	&= \frac{e}{h} \sum_\beta \int_{-\infty}^{+\infty} dE\ \mathcal T_{\alpha\beta}(E) \left[f_\alpha(E)-f_\beta(E)\right] ,
\label{totalcurrent0}
\end{align}
where the Greek letters label the terminals $\alpha=1,\cdots,\Lambda$.
The terminals, usually modeled by semi-infinite leads, are in contact with electronic reservoirs in thermal equilibrium and inject into the system electrons following a Fermi-Dirac distribution $f_\alpha(E)=[1+e^{(E-\mu_\alpha)/k_BT}]^{-1}$, where $\mu_\alpha=\mu_0+eV_\alpha$, $\mu_0$ is the equilibrium chemical potential and $V_\alpha$ is the voltage applied to the $\alpha$-terminal. For the sake of simplicity, we consider that all electronic reservoirs have the same temperature $T$.

The transmission $\mathcal T_{\alpha\beta} (E)$ is given by \cite{MeirWingreen1992}
\begin{align}
\label{eq:transmission}
\mathcal T_{\alpha\beta} (E)= {\rm tr} \left[ \mathbf \Gamma_\alpha(E) \mathbf G^r(E) \mathbf \Gamma_\beta(E) \mathbf G^a(E) \right]
\end{align}
where $\mathbf G^r=\left(\mathbf G^a\right)^\dagger$ is the retarded Green's function of the full system,
while $\mathbf \Gamma_\beta$ represents the decay width of the lead corresponding to the $\beta$ terminal.
For convenience we express $\mathbf G^r$ and $\mathbf \Gamma_\beta$ in a local basis 
representation that renders simple expressions for the local currents. Here we consider a Wannier basis with the states as labeled by $\nu=(i,\ell, \sigma)$, where the corresponding wave function is centered at the lattice atom $i$ and has orbital and spin quantum numbers given by $(\ell, \sigma)$. 
In this representation, $\mathbf G^r$ has the dimension of the number of Wannier states in the central region, while the dimension of $\mathbf \Gamma_\beta$ is the number of states corresponding to sites at the $\beta$-terminal central-region interface.
The decay width matrix reads
\begin{align}
\mathbf	\Gamma_\alpha = -2\,\text{Im}\left(\mathbf \Sigma_\alpha^r\right),
\end{align}
where $\mathbf \Sigma_\alpha^r$ is the retarded embedding self-energy, namely
\begin{align}
\mathbf{\Sigma}_\alpha^r = \mathbf V_{C \alpha} \mathbf G^r_\alpha \mathbf V^{}_{\alpha C},
\end{align}
$\mathbf V_{C \alpha}=\mathbf V_{\alpha C}^\dagger$ 
gives the coupling matrix elements between the terminal $\alpha$ and the central region $C$,
and $\mathbf G^r_\alpha$ is a contact Green's function that casts the electron dynamics in the leads.
The latter can be calculated by several methods \cite{LopezSancho1985, MacKinnon1985, Rocha2006, Wimmer2009thesis}.

In linear approximation, where $f_\alpha(E)=f_0(E)+(-\partial f_0/\partial E)eV_\alpha$ and $f_0(E)$ is the equilibrium Fermi-Dirac distribution, one obtains the familiar result \cite{Buttiker1986}
\begin{align}
I_\alpha =-\sum_{\beta=1}^{\Lambda} \mathcal G_{\alpha\beta} V_\beta = \sum_{\beta=1}^{\Lambda} \mathcal G_{\alpha\beta} \left(V_\alpha-V_\beta \right),
\label{current}
\end{align}
where $\mathcal G_{\alpha \beta}$ is the Landauer conductance given by
\begin{align}
\mathcal G_{\alpha \beta} = \frac{e^2}{h}\int_{-\infty}^{\infty} dE \left(-\frac{\partial f_0}{\partial E} \right) \mathcal T_{\alpha\beta}(E).
\label{conductance0}
\end{align}
Here $\mathcal T_{\alpha\beta}(E)$ accounts for the spin degrees of freedom and, hence, there is no spin degeneracy factor in Eq.~\eqref{conductance0}.
As standard, we chose the sign of $I_\alpha$, Eqs.~(\ref{totalcurrent0}) and (\ref{current}), to ensure a positive current from terminal $\beta$ to terminal $\alpha$ when $V_\beta>V_\alpha$ or $\mu_\beta>\mu_\alpha$.
See Appendix~\ref{appendix:lbequations} for details.

Let us now calculate the local currents using NEGF. 
We begin by considering the number operator $N_{\nu}(t) = d_{\nu}^\dagger(t) d^{}_{\nu}(t)$, where 
$d_{\nu}^\dagger$ [$d^{}_{\nu}(t)$] creates [annihilates] an electron at the 
local basis state $\nu$.
The local electron flow 
is associated to the expectation value of the temporal rate of change of the number operator, namely, 
$\langle \dot N_{\nu}(t)\rangle$. The average is taken over the grand canonical ensemble, namely, $\langle \cdots \rangle = \text{Tr}[e^{-\beta H}\cdots]/Z$, where $Z=\text{Tr}[e^{-\beta H}]$ is the partition function and $\beta = 1/k_BT$
as standard \cite{Stefanucci2013}.

We calculate $\langle \dot N_{\nu}(t)\rangle$ using the equations-of-motion method, in a similar way as it is done for the total electronic current \cite{Jauho2008,Lima2013}, and obtain
\begin{align}
	\langle \dot{N}_{\nu}(t)\rangle
	=  \int_{-\infty}^{+\infty} &\frac{dE}{h}  \bigg[ 
		\mathbf G^{<} \mathbf H	- \mathbf H	\mathbf G^{<} \nonumber\\
		& \!\!\!\!+\sum_\alpha 
		\left(\mathbf G^{<}_{C\alpha} \mathbf V_{\alpha C} - \mathbf V_{C\alpha} \mathbf G^{<}_{\alpha C}\right) \bigg]_{\nu,\nu},
\label{current3_local}
\end{align}
where $\mathbf G^<$ is the lesser Green's function of sites in the central region and $\mathbf G^<_{\alpha C}$ is hybrid lesser Green's function containing propagation information between the $\alpha$ lead and the central region. 

Recalling the identities $\mathbf H	\mathbf G^{<} = -\left(\mathbf G^{<} \mathbf H\right)^\dagger$ and $\mathbf V^{}_{C\alpha} \mathbf G^{<}_{\alpha C} = -\left(\mathbf G^{<}_{C\alpha} \mathbf V^{}_{\alpha C}\right)^\dagger$ we simplify the diagonal matrix element in Eq.~(\ref{current3_local}) to
\begin{align}
	\langle \dot N_{\nu}(t)\rangle
	&=  \frac{2}{h}\int_{-\infty}^{+\infty} \!\!{dE} \,{\rm Re}\bigg[
		\mathbf G^{<} \mathbf H	+ \sum_\alpha \mathbf G^{<}_{C\alpha} \mathbf V^{}_{\alpha C} \bigg]_{\nu,\nu}.	
\label{current4_local}
\end{align}

It is straightforward to demonstrate that 
\footnote
{
Using the standard NEGF relations 
$\mathbf G^<_{C\alpha} = 
\mathbf G^r \mathbf V_{C\alpha} \mathbf G^<_{\alpha} + 
\mathbf G^< \mathbf V_{C\alpha} \mathbf G^a_{\alpha} $
and
$\mathbf G^< = \mathbf G^r \mathbf \Sigma^< \mathbf G^a $,
one writes
$ 
\sum_\alpha \mathbf G^<_{C\alpha} \mathbf V_{\alpha C} 
= \mathbf G^r \mathbf \Sigma^< +\mathbf G^< \mathbf \Sigma^a 
= \mathbf G^< \left[ \left(\mathbf G^a\right)^{-1}  + \mathbf \Sigma^a \right]
= \mathbf G^< \left(\mathbf E -\mathbf H \right).
$
Analogously,
$
\sum_\alpha \mathbf V_{C\alpha} \mathbf G^<_{\alpha C} 
= \left(\mathbf E -\mathbf H \right)\mathbf G^<
$
}
\begin{align}
	\mathbf G^{<} \mathbf H	- \mathbf H	\mathbf G^{<}
	+\sum_\alpha 
		\left(\mathbf G^{<}_{C\alpha} \mathbf V_{\alpha C} - \mathbf V_{C\alpha} \mathbf G^{<}_{\alpha C}\right) = 0,
\end{align}
which explicitly shows that in steady state the average number of particles in any site is constant. 

In order to calculate the current flowing between any pair of states connected by the model Hamiltonian we write Eq.~(\ref{current4_local}) as
\begin{align}
	\left\langle \dot q_{\nu}(t)\right\rangle \equiv e\left\langle \dot N_{\nu}(t)\right\rangle = - \sum_{\nu'} \widetilde{I}_{\nu'\nu},
\label{localcurrent0}
\end{align}
where $\left\langle \dot q_{\nu}(t)\right\rangle$ is the rate of change of the charge in the state $\nu$ and ${\widetilde{I}}_{\nu'\nu}$ is the charge current flowing from the state $\nu$ to $\nu'$.

By comparing Eqs.~\eqref{current4_local} and \eqref{localcurrent0}, the local electronic current $\widetilde{I}_{\nu'\nu}$ can be defined as
\begin{align}
	\widetilde{I}_{\nu'\nu}
	&\equiv  -\frac{2e}{h}\!\int_{-\infty}^{+\infty} \!\!dE\ \text{Re}\bigg[ 
		 G^{<}_{\nu\nu'}  H^{}_{\nu'\nu}	+ \sum_\alpha  G^{<}_{C\alpha,\nu\nu'}  V^{}_{\alpha C,\nu'\nu} \bigg] \nonumber\\
						&= \widetilde{I}_{CC,\nu'\nu} + \sum_\alpha \widetilde{I}_{\alpha C,\nu'\nu}.
\label{current5_local}
\end{align}
Here, the subscripts $CC$ and $\alpha C$ denote the partition to which the sites corresponding to $\nu'$ and $\nu$ belong (in that order).
Thus, $H_{\nu\nu'}\neq 0$ (or $V_{\nu\nu'}\neq 0$) is a necessary condition for a nonvanishing local or bond current between the states $\nu$ and $\nu'$. 
The sum of all the bond currents across the interface between the system and the $\alpha$ lead yields 
\begin{align}
	\sum_{\nu'\nu} \widetilde{I}_{\alpha C,\nu'\nu}
	&= 
	-\frac{e}{h}\int_{-\infty}^{+\infty} \!\!dE\  
	{\rm Tr}\left[ \mathbf G^{<}_{C\alpha}  \mathbf V^{}_{\alpha C} - \mathbf V^{}_{C\alpha} \mathbf G^{<}_{\alpha C}\right],
\end{align}
which, due to charge conservation, is identical to $I_\alpha$, the total electronic current at terminal $\alpha$, given by Eq.~(\ref{totalcurrent0}) \cite{Lima2013}.
 
Let us assume that the bonds of interest are located sufficiently far from any lead interface so that all $V_{\alpha C,\nu'\nu}$ matrix elements are identically zero. Hence
\begin{align}
	\widetilde{I}_{\nu'\nu} = \widetilde{I}_{CC,\nu'\nu}
	&=  -\frac{e}{h}\int_{-\infty}^{+\infty} dE\  
	2{\rm Re}\!\left( G^{<}_{\nu\nu'}  H^{}_{\nu'\nu} \right).			
				\label{localcc}
\end{align}
Using the Keldysh equation \cite{Jauho2008} written as
\begin{align}
	\mathbf G^< = \mathbf G^r \mathbf \Sigma^< \mathbf G^a = \sum_\alpha if_\alpha\mathbf G^r \mathbf \Gamma_\alpha \mathbf G^a,
	\label{keldysh1}
\end{align}
Eq.~(\ref{localcc}) becomes
\begin{align}
	\widetilde{I}_{\nu'\nu}
	&=  \frac{e}{h}\sum_\alpha \int_{-\infty}^{+\infty} dE\ f_\alpha(E) \widetilde{\mathcal T}_{\nu'\nu}^{\alpha}(E),
\label{localcc2}
\end{align}
where
\begin{align}
	\widetilde{\mathcal T}_{\nu'\nu}^{\alpha}(E) \equiv
	2\,\text{Im}\left[ \left(\mathbf G^r \mathbf \Gamma_\alpha \mathbf G^a\right)_{\nu\nu'}  H^{}_{\nu'\nu} \right].
				\label{localtcc}
\end{align}
Recalling that $\mathbf G^r \mathbf \Gamma_\alpha \mathbf G^a$ is Hermitian, 
it is straightforward to show that 
$\widetilde{\mathcal T}_{\nu\nu'}^{\alpha}=-\widetilde{\mathcal T}_{\nu'\nu}^{\alpha}$.

Expanding $f_\alpha(E)$ up to linear order in the terminal applied voltages $\{V_\beta\}$, we write
the local current as 
\begin{align}
		\widetilde{I}_{\nu'\nu}^{}		&= \widetilde{I}_{\nu'\nu}^{\rm neq}  + \widetilde{I}_{\nu'\nu}^{\rm eq},
		\label{localcurrent}
\end{align}
where the nonequilibrium component reads
\begin{equation}
		\widetilde{I}_{\nu'\nu}^{\rm neq}\equiv  \sum_\alpha \widetilde{\mathcal G}_{\nu'\nu}^{\alpha} V_\alpha
		\label{iklneq}
\end{equation}
with a local conductance
\begin{equation}
		\widetilde{\mathcal G}_{\nu'\nu}^{\alpha} \equiv \frac{e^2}{h}\int_{-\infty}^{+\infty} dE\ \left(-\frac{\partial f_0}{\partial E}\right) \widetilde{\mathcal T}_{\nu'\nu}^{\alpha}(E).
		\label{conductancelocal}
\end{equation}
In turn, the equilibrium local charge current is given by
\begin{align}
		\widetilde{I}_{\nu'\nu}^{\rm eq}\equiv  \frac{e}{h}\int_{-\infty}^{+\infty} dE\ f_0(E) \left[\sum_\alpha \widetilde{\mathcal T}_{\nu'\nu}^{\alpha}(E)\right].
\label{ikleq}
\end{align}
Thus, for a finite bias, quantum imaging current measurements capture both the nonequilibrium and equilibrium contributions to the local currents to $\widetilde{{\bf I}}$. 

Let us project $\widetilde{I}_{\nu'\nu}^{}$ to real space. Recalling that  $\nu = (i, \ell, \sigma)$ and the atomic position can be represented by the index $i$, one can define the ``bond'' charge current as  
\begin{equation}
\widetilde I_{i'i} \equiv \sum_{\ell,\ell',\sigma, \sigma'} \widetilde{I}_{\nu'\nu}^{}.
\end{equation}
Similarly, one can define a bond spin current \cite{Nikolic2006}
\begin{align}
\widetilde I_{i'i}^s \equiv \frac{\hbar}{2e} \sum_{\ell,\ell'} \left[\widetilde{I}_{(i',\ell',\uparrow),(i,\ell,\uparrow)} - 
\widetilde{I}_{(i',\ell',\downarrow),(i,\ell,\downarrow)} \right],
\label{eq:I_spin_local}
\end{align}
where $\hbar/2e$ converts the units of a charge current into a spin one.  
We stress that Eq.~\eqref{eq:I_spin_local} applies to system Hamiltonians that do not couple spin-up with spin-down electrons, which are of interest for standard spin Hall models. In this case the spin source term vanishes and the spin currents are conserved, as nicely discussed in Ref.~\cite{Nikolic2006}. 
In general, local spin currents are present in systems with a strong spin-orbit interaction \cite{Rashba2003,Souma2005,Nikolic2006,Sousa2021}. 
By applying suitable projection schemes to $\widetilde{I}_{\nu'\nu}$, one can also write expressions for local valley currents \cite{Settnes2016,Stegmann2018} and orbital currents \cite{Bernevig2005,Cysne2021}. 

Note that, similarly to the total currents $I_\alpha$, the local nonequilibrium current $\widetilde{{\bf I}}^{\rm neq}$ is dominated by the Fermi surface states, that is, it involves an energy integration over a typically small energy window of width $\sim$$k_BT$ around $\mu_0$, 
In distinction, the evaluation of the local equilibrium currents $\widetilde{{\bf I}}^{\rm eq}$ requires an energy integration of the local transmissions 
over all system occupied states. This makes it very difficult to accurately compute $\widetilde{{\bf I}}^{\rm eq}$.
We discuss this issue in detail in Sec.~\ref{sec:QH_numerics}.

In the zero-bias limit, $\{V_\alpha\} = 0$, Eqs.~(\ref{current}) and (\ref{localcurrent}) yield
\begin{align}
{I}_\alpha = 0 \qquad \mbox{and} \qquad 
\widetilde{I}_{\nu'\nu} = 	\widetilde{I}_{\nu'\nu}^{\rm eq}.
\end{align}
These relations imply that if $\sum_\alpha \widetilde{\mathcal T}_{\nu'\nu}^{\alpha}(E) \neq 0$ it is possible to have a finite local electronic current in the central region, $\widetilde{I}_{\nu'\nu}^{\rm eq} \neq 0$, even in the absence of a bias voltage, but there is no net current flowing through the contacts ${I}_\alpha = 0$. The latter implies that there is no entropy production \cite{Bruch2018}, as required by equilibrium processes. 
Equation (\ref{localcc2}) indicates that each lead $\alpha$ contributes by injecting electrons in the system for all energies up to its chemical potential $\mu_\alpha$. 
Therefore, in the zero-bias case ($\{V_\alpha\} = 0$) all the leads inject electrons into the system at all energies up to the equilibrium chemical potential $\mu_0$.
In this sense, the equilibrium currents can be viewed as a property of the system many-electron ground state.

We can infer an important property of the local equilibrium currents using the charge conservation and a control cross section, such as the one indicated in Fig.~\ref{fig:hall_bars_setup_A}. 
The continuity equation demands that the local current $\widetilde{{\bf I}}^{\rm eq}$ integrated along a closed domain is zero.
By taking a cross section that contains a terminal $\alpha$ and recalling that at equilibrium $I_\alpha=0$, we show that the integral of $\widetilde{{\bf I}}^{\rm eq}$ over any system cross section is zero.

In what follows we explore situations where the equilibrium steady-state currents injected by all terminals do not cancel each other and there is a nonvanishing local current in the absence of bias voltage.

\section{Toy model}
\label{sec:toymodel}

\begin{figure}
	\centering
		\includegraphics[width=1.00\columnwidth]{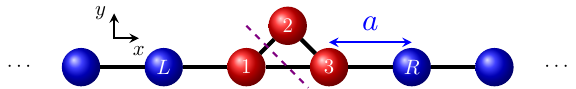}
	\caption{Sketch of the toy model system: Three sites in the central region attached to two semi-infinite leads at left ($L$) and right ($R$).
	The dashed line represents the cross section used to evaluate the net local equilibrium current in the main text.
	}
	\label{fig:toymodel}
\end{figure}

Let us illustrate some of the main features of the charge equilibrium currents using an analytically solvable (nontopological) toy model. 
We consider a central region consisting of 3 sites coupled to semi-infinite linear chains, see Fig.~\ref{fig:toymodel}.
We assume that the electrons are described by a nearest-neighbor single-orbital tight-binding Hamiltonian with hopping matrix elements $-t_{kk'}$ with no spin dependence. 
The central region sites lie in the $xy$ plane at $\mathbf r_1 = (0,0)$, $\mathbf r_2 = (a/2,a/2)$, $\mathbf r_3 = (a,0)$. 
We account for a constant magnetic field $\mathbf B = B\hat {\bf z}$ using the Peierls substitution, namely,
$t_{kk'} \rightarrow t_{kk'}e^{i\varphi_{kk'}}$, where  
$\varphi_{kk'} = \frac{eB}{2\hbar} \left(x_{k'}-x_k\right) \left(y_{k'}+y_k\right)$ \cite{Lewenkopf2013}.
Thus,
\begin{align}
	\varphi_{12} = \varphi_{23} = \phi/2
	\quad \text{and} \quad
	\varphi_{13} = 0,										
\end{align}
where $\phi \equiv 2\pi \Phi/\Phi_0$ is proportional to the ratio between the magnetic flux $\Phi = Ba^2/4$ enclosed by the central region triangular ``loop" (see Fig.~\ref{fig:toymodel}) and the magnetic flux quantum $\Phi_0 = h/e$. 

Assuming that the sites $1$ and $3$ are attached to leads with self-energies that, in the wide band limit, can be approximated by $\Sigma_1^r=-i\Gamma_1/2$ and $\Sigma_3^r=-i\Gamma_3/2$, we write $\mathbf G^r(E)=(E\mathbf 1 - \mathbf H - \mathbf \Sigma^r)^{-1}$ as
\begin{align}
  \mathbf G^r(E) &=
	\left(\begin{array}{ccc}
		E+i\Gamma_1/2				& t_{12}e^{i\phi/2}		& t_{13}            \\
		t_{12}e^{-i\phi/2}	& E										& t_{23}e^{i\phi/2} \\
		t_{13}							& t_{23}e^{-i\phi/2}	& E+i\Gamma_3/2
	\end{array} \right)^{-1}.
	\label{ehstoymodel}
\end{align}

{
It is instructive to verify charge conservation by calculating the net electronic transmission through the central region sites.
Considering the injection from the left lead $L$, the net electronic transmissions read
\begin{align}
	\widetilde{\mathcal T}^{L}_{12}+\widetilde{\mathcal T}^{L}_{13} &=-\frac{\Gamma_1\Gamma_3}{|D|^2} ( \tau_4-E\tau_3\cos\phi+ E^2\tau_2 ),\label{localt1}\\
	\widetilde{\mathcal T}^{L}_{21}+\widetilde{\mathcal T}^{L}_{23} &=0, \label{localt2}\\
	\widetilde{\mathcal T}^{L}_{31}+\widetilde{\mathcal T}^{L}_{32} &=\frac{\Gamma_1\Gamma_3}{|D|^2} ( \tau_4-E\tau_3\cos\phi+ E^2\tau_2 ),\label{localt3}
\end{align}
where $D = {\rm det}[ ({\bf G}^r)^{-1}], \tau_2\equiv t_{13}^{2}$, $\tau_3\equiv 2t_{12}t_{13}t_{23}$, and $\tau_4\equiv t_{23}^{2}t_{12}^{2}$. The derivation of the expressions for the transmissions are presented in  Appendix~\ref{app:gfsandtransmissions} and the sign convention is discussed in Appendix~\ref{app:toymodel-sign-convention}. 
The net electronic transmission at site $2$, given by Eq.~(\ref{localt2}), vanishes due to charge conservation, that is, the local transmission from site $1$ to site $2$ equals the local transmission from site $2$ to site $3$.
In contrast, the electronic transmissions given by Eqs.~(\ref{localt1}) and (\ref{localt3}), respectively, do not vanish since {they do not account for} the transmissions through the bonds with the leads, i.e., electrons entering site $1$ through the left lead $L$ and leaving site $3$ through the right lead $R$.
Moreover, assuming injection from the left lead $L$, the total transmission leaving site $1$, Eq.~(\ref{localt1}), must equal the transmission entering site $3$, Eq.~(\ref{localt3}).
}

{
The total transmission between the left and the right leads reads
\begin{align}
\mathcal T_{RL} 
= \frac{\Gamma_1\Gamma_3}{|D|^2} ( \tau_4-E\tau_3\cos\phi+ E^2\tau_2 )
\label{trltoymodel}
\end{align}
This ensures charge conservation 
for site $1$ since $\widetilde{\mathcal T}^{L}_{1L}+\widetilde{\mathcal T}^{L}_{12}+\widetilde{\mathcal T}^{L}_{13}=0$ and 
for site $3$ since $\widetilde{\mathcal T}^{L}_{3R}+\widetilde{\mathcal T}^{L}_{31}+\widetilde{\mathcal T}^{L}_{32}=0$. 
Here we {have} used that $\widetilde{\mathcal T}^{L}_{1L}={\mathcal T}_{RL}$ and $\widetilde{\mathcal T}^{L}_{3R}=-{\mathcal T}_{RL}$. 
The above arguments are also valid assuming injection from the right lead.
An analogous calculation {shows} that
$\widetilde{\mathcal T}^{R}_{3R}+\widetilde{\mathcal T}^{R}_{31}+\widetilde{\mathcal T}^{R}_{32}=0$ and
$\widetilde{\mathcal T}^{R}_{1L}+\widetilde{\mathcal T}^{R}_{12}+\widetilde{\mathcal T}^{R}_{13}=0$.
}

We are now ready to analyze the equilibrium current. Let us consider, for instance, the bond current flowing from site 2 to site 1, namely
\begin{equation}
\widetilde{I}_{12}^{\rm eq} = \frac{e}{h} \int_{-\infty}^\infty dE f_0(E) \left[ \widetilde{\mathcal T}^{L}_{12}(E) +\widetilde{\mathcal T}^{R}_{12}(E)\right]
\label{i12eq}
\end{equation}
with the sum of local transmissions given by 
\begin{align}
\widetilde{\mathcal T}^{L}_{12}&+\widetilde{\mathcal T}^{R}_{12} = 
-\frac{2}{|D|^2}t_{12}t_{23}t_{13} \times \nonumber\\
& \times  \left[ (\Gamma_1+\Gamma_3) E^2 - (\Gamma_1 t_{23}^2 + t_{12}^2 \Gamma_3)  \right] \sin\phi.
\label{3x3teq}
\end{align}

Interestingly, $\widetilde{\mathcal T}^{L}_{12}+\widetilde{\mathcal T}^{R}_{12}$ (and thus $\widetilde{I}_{12}^{\rm eq}$)
vanishes in three situations: 
(i) if one breaks the loop by turning off any of the hopping matrix elements $t_{12}$, $t_{23}$ or $t_{13}$, 
(ii) if there is no magnetic field ($\phi=0$), and 
(iii) if the system is detached from the leads ($\Gamma_1=\Gamma_3=0$).

The presence of a magnetic field breaks time reversal symmetry inducing a preferential electronic flow through one of the two system branches, $1$-$3$ or $1$-$2$-$3$, depending on the injection direction. 
This causes a current imbalance between the branches leading to a nonvanishing equilibrium electronic current given by Eq.~(\ref{i12eq}). Conversely, once the loop is broken by disconnecting one of the bonds, the electronic current flows through a single branch irrespective of the injection direction and there is no local current in the equilibrium.

The lack of equilibrium currents when the system is detached from the leads is somehow surprising in view of the vast  theoretical  \cite{Buttiker1983, Altshuler1991, MullerGroeling1993} and experimental \cite{Levy1990,Bluhm2009,Bleszynski-Jayich2009} literature on persistent (equilibrium) currents in isolated mesoscopic rings.
Unfortunately, the Landauer-B\"uttiker approach is unable to address isolated systems.
Periodic boundary conditions play a key role to explain equilibrium persistent currents, while in our formulation the scattering states necessarily involve both the central region and the leads and, thus, cannot be reduced to an isolated system setup
\footnote{As discussed in the forthcoming sections, the underlying physics of the IQH and QSH regimes is rather different and these kinds of systems are nicely described by the Landauer-Buettiker approach.}.

The simplicity of the model allows us to analytically show that 
\begin{align}
 \widetilde{\mathcal T}^{L}_{12}+\widetilde{\mathcal T}^{R}_{12} 
=\widetilde{\mathcal T}^{L}_{31}+\widetilde{\mathcal T}^{R}_{31} 
=\widetilde{\mathcal T}^{L}_{23}+\widetilde{\mathcal T}^{R}_{23}.
\label{eq:Tsums_toymodel}
\end{align}
The above expressions highlight two important properties of equilibrium currents: 

(i) {\sl There is no net equilibrium current leaving or entering the system:} 
The total equilibrium transmission from $2$ to $1$, given by the sum in Eq.~(\ref{i12eq}), namely $\widetilde{\mathcal T}^{L}_{12}+\widetilde{\mathcal T}^{R}_{12}$, equals the total equilibrium transmission from $1$ to $3$ given by $\widetilde{\mathcal T}^{L}_{31}+\widetilde{\mathcal T}^{R}_{31}$. 
By invoking charge conservation and by accounting for the injection of both left and right terminals, we find that there is no net equilibrium transmission between $1$ and $L$. 
The same happens for the net equilibrium transmission between the site $3$ and the right reservoir $R$.
In summary, both terminals inject the same equilibrium current into the system leading to a zero net equilibrium electronic current, that is, $\widetilde{I}^{\rm eq}_{1L} = \widetilde{I}^{\rm eq}_{3R} =0$. A nonvanishing net electronic current at the terminals requires a voltage bias. In this case, Eq.~(\ref{iklneq}) does not vanish and a nonequilibrium electronic current sets in.

(ii) {\sl The equilibrium current integrated over any system cross section is identically zero.}
This is nicely seen by analyzing the equilibrium current $\widetilde{I}^{\rm eq}_{\rm sec}$ flowing through the section 
defined by the dashed line in Fig.~\ref{fig:toymodel}, namely,
\begin{equation}
		\widetilde{I}_{\rm sec}^{\rm eq}\equiv  \frac{e}{h}\int_{-\infty}^{+\infty} dE\ f_0(E)
		 \left[\widetilde{\mathcal T}^{L}_{12}+\widetilde{\mathcal T}^{R}_{12} 
                  +\widetilde{\mathcal T}^{L}_{13}+\widetilde{\mathcal T}^{R}_{13} \right].
\label{iseceq}
\end{equation}
Recalling that $\widetilde{\mathcal T}^\alpha_{ij}=-\widetilde{\mathcal T}^\alpha_{ji}$ and Eq.~\eqref{eq:Tsums_toymodel}, 
it immediately follows that $\widetilde{I}^{\rm eq}_{\rm sec} =0$.

\section{Local currents in the quantum Hall regime}
\label{sec:IQH}

In this section we study the equilibrium and nonequilibrium local currents in the IQH regime \cite{Prange1990} for a multiprobe setup.
First, we discuss general qualitative aspects of the transport properties using the Landauer-B\"uttiker approach.
Next, we calculate the local and total currents using the formalism presented in Sec.~\ref{sec:multiprobe}
considering a graphene  system as a case in point.

Let us consider $6$-terminal setup, as the one sketched in Fig.~\ref{fig:hall_bars_setup_A}. 
{We address the situation} where an applied bias voltage $V_{\rm bias}$ between terminals $\alpha=1$ and $\beta=6$ drives an electronic current $I$ from terminal $1$ ($I_1=I$) to terminal $6$ ($I_6=-I$).  
For simplicity, we set $V_1=V_{\rm bias}$ and assume the terminal $6$ as grounded, $V_6=0$.
We consider terminals $2$ through $5$ as voltage probes, namely, $I_2=I_3=I_4=I_5=0$.
The current $I$ and the voltages $V_2$, $V_3$, $V_4$ and $V_5$ are determined by the applied bias $V_{\rm bias}$.

\subsection{General discussion}
\label{sec:QH_general}

Here we address qualitatively the charge transport properties of a Hall bar system in the IQH regime \cite{Prange1990,Datta1995}  paying particular attention to equilibrium currents.

We consider that the Hall bar, see Fig.~\ref{fig:hall_bars_setup_A}, is subjected to a perpendicular magnetic field $\mathbf B = B \mathbf{\hat z}$ that is sufficiently strong to give rise to quantized edge states. 
The electrons injected at the terminal $\alpha=1$ along the $x$ direction are deflected towards their left-hand side flowing to the nearest terminal $\beta=2$ with edge states propagating along the $y$ direction.

\begin{figure}[tbp]
	\centering
		\includegraphics[width=0.49\columnwidth]{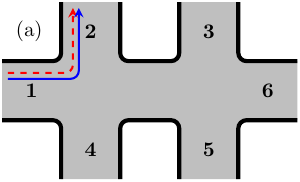}
		\includegraphics[width=0.49\columnwidth]{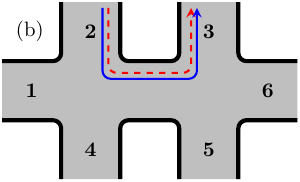}
		\includegraphics[width=0.49\columnwidth]{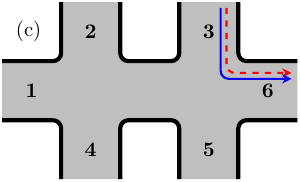}
		\includegraphics[width=0.49\columnwidth]{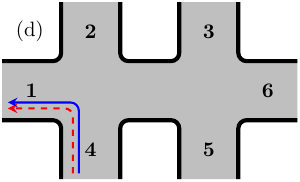}
		\includegraphics[width=0.49\columnwidth]{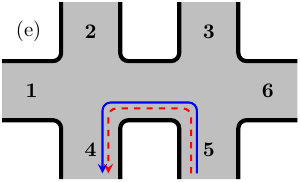}
		\includegraphics[width=0.49\columnwidth]{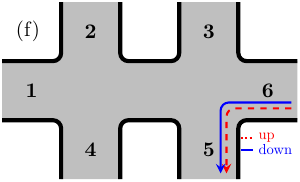}
	\caption{
	Sketches of local transmissions/conductances for a system in the IQH regime. 
	Panels (a) to (f) correspond to an electronic injection from a single terminal $\alpha=1$ to $6$, respectively. 
	Spin dependent process are neglected and, thus, spin-up (dashed red) and spin-down (solid blue) states are degenerate.
	}
		\label{fig:hall_bar_QH_local_t}
\end{figure}

Figure~\ref{fig:hall_bar_QH_local_t} shows the local transmissions for all possible single terminal injection processes.
Due to the strong magnetic field electrons are transmitted by edge states, the conductance is quantized, and the current is chiral \cite{Datta1995}.
In this section we neglect spin-dependent processes and, thus, the edge states are spin degenerate.

For simplicity, we consider the zero-temperature limit and, further, assume that $E_F$ lies between the $N_{\rm L}$th and $(N_{\rm L}+1)$th Landau level energies.
In this situation, there are $N=N_{\rm L}$ propagating channels per spin at any of the system edges.
The conductance matrix, given by Eq.~(\ref{conductance0}), reads 
\begin{align}
	\boldsymbol{\mathcal{G}} = 
	2N\frac{e^2}{h}
	\left(\begin{array}{cccccc}
-1 & 0 & 0 & 1 & 0 & 0\\
 1 &-1 & 0 & 0 & 0 & 0\\
 0 & 1 &-1 & 0 & 0 & 0\\
 0 & 0 & 0 &-1 & 1 & 0\\
 0 & 0 & 0 & 0 &-1 & 1\\
 0 & 0 & 1 & 0 & 0 &-1
\end{array}\right).
\label{conductanceqh}
\end{align}
The negative diagonal conductance values manifestly enforces current conservation.
Current conservation and gauge invariance imply that
$\sum_\alpha \mathcal G_{\alpha\beta} =\sum_\beta \mathcal G_{\alpha\beta} =0$ \cite{Buttiker1986,Ihn2010}.

Using Eqs.~(\ref{current}) and (\ref{conductanceqh}) we calculate the unknown voltages $\{V_\alpha\}$ for an applied bias voltage $V_{\rm bias}$ between terminals $1$ and $6$ that drives an electronic current $I$, see Fig.~\ref{fig:hall_bars_setup_A}.
We obtain
\begin{align}
\hspace{-0.04cm}
\left(
 V_1,  V_2,  V_3,  V_4,  V_5,  V_6 \right)
=
\left(
 1, 1, 1,  0, 0, 0\right)\frac{h}{2Ne^2}I,
\label{voltagesqh}
\end{align}
where $I=(2Ne^2/h)V_{\rm bias}$.
As expected, the ``top'' terminals $2$ and $3$ are in equilibrium with the source terminal $1$, the ``bottom'' terminals $4$ and $5$ are in equilibrium with the drain terminal $6$, and there is no voltage drop through the propagation direction ($V_2=V_3, V_4=V_5$) \cite{Datta1995}.

The longitudinal resistance \cite{Buttiker1986,Datta1995,Ihn2010} $R_{xx} \equiv R_{16,45} \equiv \left|V_2-V_3 \right|/I  = R_{16,23} = 0$ vanishes, while the transverse resistance is $R_{xy}  \equiv R_{16,35} \equiv \left|V_3-V_5 \right|/I = R_{16,24} = h/2Ne^2$.
This is the standard B\"uttiker picture to describe IQH transport measurements \cite{Buttiker1986}.

Additionally, we calculate the contact resistance $R_c$ by subtracting the longitudinal resistance $R_{xx}$ from $R_{16,16}$, namely $R_{c} \equiv R_{16,16} - R_{xx} = h/2Ne^2$.
{This shows that} the contact resistance $R_c$ includes contributions from the system coupling to both terminals $1$ and $6$ and that each terminal offers a contact resistance to the established current $I$.
Thus, in the IQH regime the resistance $R_{16,16}=h/2Ne^2$ corresponds solely to the resistance generated by $2N$ propagating modes at the contacts \cite{Datta1995,Ihn2010}.

Let us use the B\"uttiker voltage probe model \cite{Buttiker1986b} to briefly discuss decoherence effects and why the longitudinal resistance vanishes in the IQH regime. The introduction of an extra terminal $\varphi$ to a $\Lambda$-terminal system can be trivially accounted for by writing Eq.~\eqref{current} as
\begin{align}
I_\alpha &= \mathcal G_{\alpha\varphi} \left(V_\alpha-V_\varphi \right) + \sum_{\beta=1}^{\Lambda} \mathcal G_{\alpha\beta} \left(V_\alpha-V_\beta \right), 
\label{currentalpha}\\
I_\varphi &= \sum_{\beta=1}^{\Lambda} \mathcal G_{\varphi\beta} \left(V_\varphi-V_\beta \right).
\label{currentvarphi}
\end{align}
Assuming that the terminal $\varphi$ is a voltage probe ($I_\varphi=0$), one can calculate $V_\varphi$ using Eq.~(\ref{currentvarphi}) and substitute the result into  Eq.~(\ref{currentalpha}) to obtain \cite{Datta1995}
\begin{align}
I_\alpha &= \sum_{\beta=1}^{\Lambda} { \mathcal G}_{\alpha\beta}^{\rm eff} \left(V_\alpha-V_\beta \right) ,
\end{align}
where the effective conductance is
\begin{align}
{\mathcal G}_{\alpha\beta}^{\rm eff} \equiv \mathcal G_{\alpha\beta} + \frac{\mathcal G_{\alpha\varphi}\mathcal G_{\varphi\beta}}{\sum_{\beta=1}^\Lambda \mathcal G_{\varphi\beta}}.
\label{conductanceeff}
\end{align}

For simplicity, let us discuss the influence of a {maximally coupled} voltage probe on a two-terminal setting ($\Lambda=2$), with ``left'' and ``right'' leads.
In the IQH regime for $N=1$, 
the conductance  reads  $\mathcal G_{LR}=\mathcal G_{RL}=2e^2/h$.
The addition of an extra terminal $\varphi$ interrupts the direct flow between terminals $L$ and $R$, leading to the nonvanishing conductance elements
$\mathcal G_{L\varphi}=\mathcal G_{\varphi R}=\mathcal G_{RL}=2e^2/h$. 
The extra terminal changes $\mathcal G_{LR}$ from $2e^2/h$ to $0$ and preserves $\mathcal G_{RL}$.
{By inserting these results in Eq.~\eqref{conductanceeff} one finds that  ${\mathcal G}_{LR}^{\rm eff}={\mathcal G}_{RL}^{\rm eff}=2e^2/h$.
Thus, the introduction of an extra terminal does not change the effective conductance of the system.

Usually a reservoir incoherently populates the system channels producing additional resistance \cite{Protogenov2013, Datta1995}. 
But in the IQH regime, 
the chiral nature of the states forces the extra terminal to collect the entire electronic flow and inject it back.
Thus, decoherence processes do not introduce momentum relaxation here, leading to the absence of longitudinal resistance.

Let us now qualitatively discuss the local conductances and currents.
Figure~\ref{fig:hall_bar_QH_local_i_neq} sketches the nonequilibrium local electronic current, Eq.~(\ref{iklneq}) and \eqref{conductancelocal}, for  the voltages given by Eq.~(\ref{voltagesqh}).
For $N=1$, the value of the local conductance at each edge is $2e^2/h$ 
and $V_1=V_2=V_3=hI/2e^2$, leading to an electronic current $I$ flowing through the system upper edges.
In contrast, since $V_4=V_5=V_6=0$, there is no electronic current flowing at the bottom edge of the Hall bar. 

\begin{figure}[htbp]
	\centering
		\includegraphics[width=0.75\columnwidth]{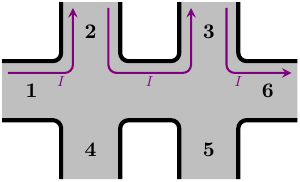}
	\caption{
		Nonequilibrium component of the local electronic current map $\widetilde{I}_{\nu'\nu}^{\rm neq}$ estimated using Eq.~(\ref{iklneq}). 
		Since the states are spin degenerate both spin components are sketched together for simplicity.
	}
		\label{fig:hall_bar_QH_local_i_neq}
\end{figure}

In turn, the local {\it equilibrium} current map is obtained from Eq.~(\ref{ikleq}).
As discussed in Sec.~\ref{sec:toymodel}, the magnetic field breaks time reversal symmetry causing a current imbalance at the central branch of the Hall bar, the sum of transmissions in Eq.~(\ref{ikleq}) does not vanish locally resulting in a clock-wise circulating electronic current.
Since Eq.~(\ref{ikleq}) involves an energy integration over all occupied states,  $\widetilde{I}_{\nu'\nu}^{\rm eq}$ contains contributions from both edge states and bulk ones. 
We discuss these features in the next subsection for a given model system.

\subsection{Local currents in Quantum Hall systems: Numerical results}
\label{sec:QH_numerics}

Let us now present a quantitative analysis of the local currents.
For that purpose we consider, as an example, a graphene system modeled by a single-orbital nearest-neighbor tight-binding Hamiltonian, namely \cite{CastroNeto2009}
\begin{align}
	H = -\sum_{\left<i,j\right>,\sigma} t_{ij} c_{i\sigma}^\dagger c^{}_{j\sigma},
	\label{hamiltonianqh}
\end{align}
where $c_{i\sigma}^\dagger$($c_{i\sigma}^{}$) is the operator that creates (annihilates) an electron with spin projection $\sigma$ at the $i$th site of a honeycomb lattice, $t_{ij}=te^{i\varphi_{ij}}$ with $t= 2.7$ eV,  $\varphi_{ij}$ is the Peierls phase discussed in Sec.~\ref{sec:toymodel} that accounts for a constant magnetic field ${\bf B}=B\hat{\bf z}$, and $\left< \cdots \right>$ restricts the summation to nearest neighbors sites. This simple Hamiltonian is very successful in describing the low-energy properties of monolayer graphene samples \cite{CastroNeto2009}.

The transport properties are calculated using the multiterminal RGF method presented in Ref.~\cite{Lima2018}. 
The Hall bar has armchair (zigzag) edges along the horizontal (vertical) direction, which is about $900$ \AA ~($530$ \AA) long. 
As standard \cite{Lewenkopf2013}, we model the contacts by using pristine graphene nanoribbons in the absence 
of a magnetic field with high doping in order to mimic the large density of states of the metallic contacts used in experiments. 
We have checked that the leads are sufficiently wide to avoid edge-to-edge interface coupling. Our simple choice for the system-contact 
interface is justified by Ref.~\cite{Santos2019}, that by using different convenient gauges (see, for instance, Ref.~\cite{Cresti2021}) 
showed that the transport properties of graphene systems in the IQH regime are almost insensitive to the degree of smoothness of the system-contact 
interface.  

The Landau levels energies $E_{N_{\rm L}}$ {are given in good approximation, as long as $\left|E_{N_L}\right|\ll t$ \cite{Lima2018}, by} $E_{N_{\rm L}}=E_1\sqrt{N_{\rm L}}$ \cite{CastroNeto2009}, where $E_1=\sqrt{3/2}a t/\ell_B$ and $\ell_B=\sqrt{\hbar/eB}=\sqrt{A_H/2\pi(\Phi/\Phi_0)}$ is the magnetic length.
Here, $\Phi \equiv BA_H$ is the magnetic flux enclosed by a single hexagon of the honeycomb lattice of area $A_H=a^2\sqrt{3}/2$, the magnetic quantum flux is $\Phi_0\equiv h/e$ and $a=2.46$ \AA.
We set 
$\Phi/\Phi_0=0.01$, 
so that $E_1=0.33t$. 

\begin{figure}[th]
	\centering
		\includegraphics[width=0.99\columnwidth]{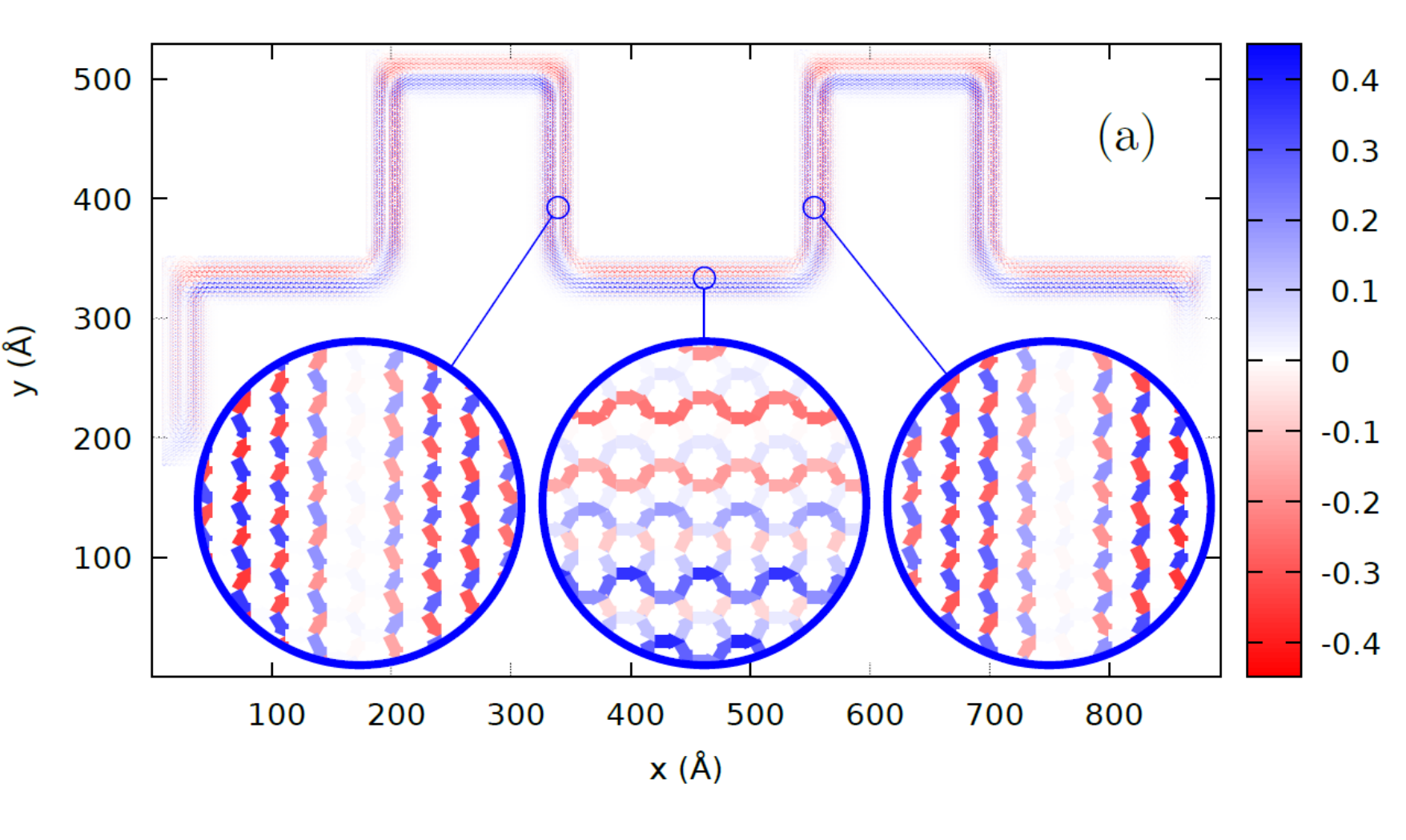} 
		\includegraphics[width=0.99\columnwidth]{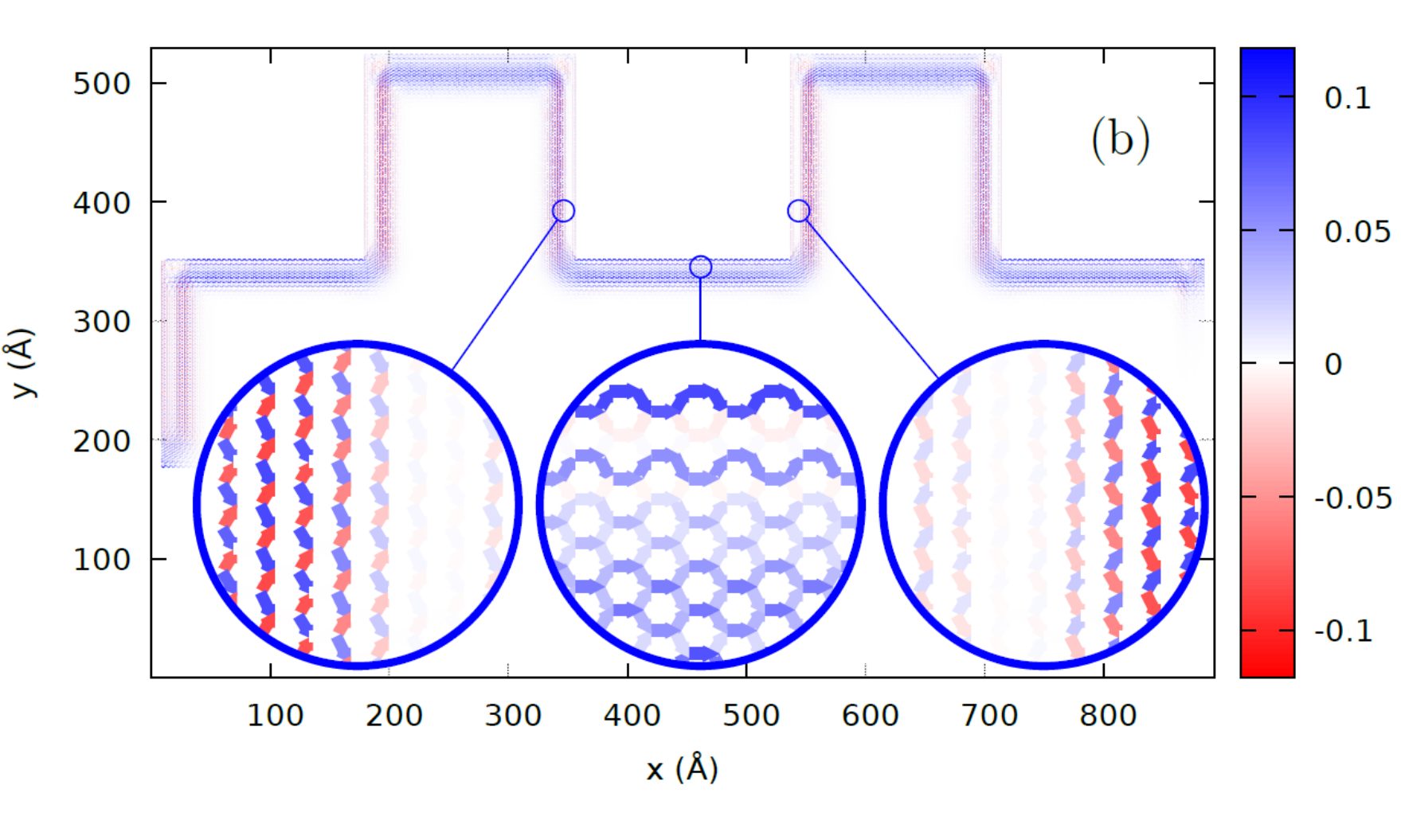} 
	\caption{
		Nonequilibrium local electronic charge current $\widetilde{\bf I}^{\rm neq}$ in the IQH regime calculated using Eq.~(\ref{iklneq}) for (a) $E_F=0.333t$ and (b) $E_F=0.4t$.
		The current is in units of $e^2V_{\rm bias}/h$ and includes both spin orientations.
		{Positive values (blue) indicate local currents flowing along the arrow directions, while negative values (red) indicate currents opposite to the arrows.}
	}
		\label{fig:IQH_neq}
\end{figure}

Figure~\ref{fig:IQH_neq} shows that the {\it nonequilibrium} local charge current at the atomistic level of a graphene Hall bar in the IQH regime for $E_F = 0.333 t$ and  $E_F = 0.4 t$, {just above the $N_{\rm L}=1$ Landau level energy,} follows the behavior  predicted in Sec.~\ref{sec:QH_general}, see Fig.~\ref{fig:hall_bar_QH_local_i_neq}. Both the intensity and the direction of the electronic current are given by the arrow directions and their color intensity. The 
honeycomb lattice has three distinct bond directions and we attribute different colors when the bond current flows towards (blue) or against (red) the arrows directions. The local electronic current has contributions from terminals $1$, $2$, and $3$ that are equilibrated at the same voltage.
Here, terminals $4$, $5$, and $6$ play no role in Eq.~(\ref{iklneq}) since their voltages vanish.
The resulting current map shows an electronic flow from source (left) to drain (right) that runs through the upper edges, as expected from the qualitative discussion of the previous subsection \cite{Buttiker1988}.

The insets of Fig.~\ref{fig:IQH_neq}(a) show that for $E_F = 0.333 t$ the electronic current has two counter propagating components due to interference effects caused by scattering at the corners formed between zigzag and armchair edges in the middle of the Hall bar.
The component flowing towards the drain lies further away from the edges while the backscattered component lies at the edge vicinity.
This simple picture is ratified by noticing that the interference effect is energy dependent.
Changes on the electronic energy $E_F$ that lead to variations of the electron group velocity, modify the interference pattern.
We find (not shown here) that this effect is stronger for $E_F$ in the vicinity of the Landau levels energies $E_{N_L}$, where the group velocity displays an enhanced dependence on $E_F$. 
We show in Fig.~\ref{fig:IQH_neq}(b) that for $E_F = 0.4 t$ the counter propagating flow of electrons disappears because the ratio between the Fermi wavelength and the distance between the corners changed.
Notice that the number of propagating channels is $3$ and the current integrated over a vertical cross section has the same value for both cases in Fig.~\ref{fig:IQH_neq}. 

\begin{figure}[ht]
	\centering
		\includegraphics[width=0.99\columnwidth]{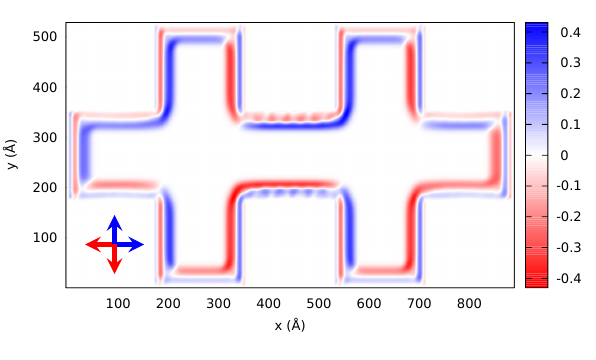} 
	\caption{
		Equilibrium local electronic charge current $\Delta \widetilde{\bf I}^{\rm eq}(\mu_1, \mu_2)$ in the IQH regime calculated using Eq.~(\ref{deltaikleq}) for $\mu_1 = 0.3t$ and $\mu_2 = 0.4t$.
		The current is in units of $et/h$ and includes both spin orientations.
		The arrows indicate the color code for the electronic propagation direction. Propagation towards the right and top directions have positive values (blue) while propagation towards left and bottom have negative values (red). 	
	}
		\label{fig:IQH_eq}
\end{figure}

We obtain the local equilibrium current by evaluating Eq.~(\ref{ikleq}) for each bond in the system.
The computation of Eq.~(\ref{ikleq}) is very time consuming since the integrand presents sharp variations with energy that require a high density of points to be resolved and the integral has to be calculated over a large energy interval, that starts at the bottom of the band and accounts for all occupied states.
More importantly, the result is hardly comparable with experiments, since it involves a precise description of the system specific delocalized states that appear between successive plateaus in the IQH regime and strongly depend on the disorder configuration.}
To circumvent those issues we propose the analysis of 
$	\Delta\widetilde{I}_{\nu'\nu}^{\rm eq}(\mu_1,\mu_2) \equiv  \widetilde{I}_{\nu'\nu}^{\rm eq}(\mu_2) - \widetilde{I}_{\nu'\nu}^{\rm eq}(\mu_1)$ instead.
At the zero-temperature limit, $\Delta\widetilde{I}_{\nu'\nu}^{\rm eq}(\mu_1,\mu_2)$, reads
\begin{align}
	\Delta\widetilde{I}_{\nu'\nu}^{\rm eq}(\mu_1,\mu_2) = \frac{e}{h}\int_{\mu_1}^{\mu_2} dE\ \left[\sum_\alpha \widetilde{\mathcal T}_{\nu'\nu}^{\alpha}(E)\right].
\label{deltaikleq}
\end{align}
This quantity can be experimentally determined by performing two distinct local current measurements with the system doped at different chemical potentials using a similar setting as that of the experiments reported in Ref.~\cite{Uri2020}. 
This simple two-measurement protocol allows a comparison with theory, since it facilitates the computation of Eq.~(\ref{ikleq}) for realistic model Hamiltonians and doping differences.

Figure~\ref{fig:IQH_eq} shows the local {\it equilibrium} current map, obtained by integrating Eq.~(\ref{ikleq}) from {$\mu_1=0.300t$ up to $\mu_2=0.400t$} to capture the transition though the Landau level $N_L=1$, that is, the transition between $1$ to $3$ propagating edge modes per spin \cite{Lima2018}. 
As predicted by the qualitative analysis of Sec.~\ref{sec:QH_general},
the equilibrium charge current flows clockwise at the system edges.
The equilibrium local current has contributions from edge states with energies above and below $E_1$.
We recall that the transmission at a particular electronic energy can exhibit an interference pattern that presents forward and backwards propagation along the edge direction due to interference effects caused by scattering at the zigzag-armchair corners as illustrated by Fig.~\ref{fig:IQH_neq}.
Interestingly the local equilibrium current, that is composed by integrating the sum of $\widetilde{\cal T}^{\alpha}_{\nu\nu'}(E)$ that captures different interference effects, presents a resulting pattern characterized by a clear clockwise electronic current at the edges and an additional counter-clockwise electronic current closer to the system center.

Figure~\ref{fig:IQH_eq} also shows that the equilibrium current flows parallel to the system-leads interfaces. 
This is a consequence of the mismatch between the system modes and the leads modes (doped at a high density of states energy, realized by shifting the energy of the electrons at the contacts by $-t$).
The current injected by a given lead enters the system through all sites at the corresponding lead-system interface.  
The magnetic field inside the central region forces the current to flow parallel to the system-lead interface towards the edge. 
The superposition of the injection by all six terminals {of the Hall bar} renders the electronic flow that is parallel to the system-leads interfaces, as seen in Fig.~\ref{fig:IQH_eq}.

We recall that the local {\sl equilibrium} electronic current integrated over any cross section of the system vanishes, 
{as shown in Sec.~\ref{sec:multiprobe}.}
We have integrated $\widetilde{\bf I}^{\rm eq}$ over several different cross sections and verified that, within numerical precision, this statement is correct.  

We conclude this section by stressing that equilibrium charge currents cannot be assessed by nonequilibrium transport measurements, since the latter inevitably involve irreversible processes, such as electron equilibration at the contacts probes.
However, since the equilibrium local currents in the IQH regime are chiral, they produce a magnetic field that can in principle be measured. 
Having obtained $\widetilde{I}^{\rm eq}_{ij}$ for a given system, the change in magnetic field $\Delta B ({\bf r})$ can be calculated using the Biot-Savart law, as discussed in Ref.~\cite{Chen2020}. Experiments can follow the reverse path, for instance, use SQUIDs to measure $\Delta B ({\bf r})$ 
\cite{Uri2020} and infer the local equilibrium current.

\section{Local currents in Quantum spin Hall systems}
\label{sec:QSH}

In this section we study the electronic and spin transport of multiterminal {two-dimensional} systems 
in the QSH regime.
As above, for simplicity we consider the zero temperature limit.
For Fermi energies within the topological gap, the nonequilibrium electronic transport in a Hall bar geometry is ruled by helical states propagating at the system edges and the conductances ${\cal G}_{\alpha\beta}$ are quantized \cite{Hasan2010,Qi2011,Bernevig2013}. 
We show that the equilibrium currents do not follow this simple picture.

\subsection{General discussion}
\label{sec:QSH_general}

The conductance matrix elements $\mathcal G_{\alpha \beta}$ of an electronic system in the QSH regime at $T=0$ in a Hall bar configuration {shown in Fig.~\ref{fig:hall_bars_setup_A} reads \cite{Protogenov2013,Mani2016}}
\begin{align}
	\boldsymbol{\mathcal{G}} = 
	\frac{e^2}{h}
	\begin{pmatrix}	
-2&1 &0 &1 &0 &0\\
1 &-2&1 &0 &0 &0\\
0 &1 &-2&0 &0 &1\\
1 &0 &0 &-2&1 &0\\
0 &0 &0 &1 &-2&1\\
0 &0 &1 &0 &1 &-2
\end{pmatrix}.
\label{conductanceqsh}
\end{align}
The quantized entries reflect the fact that in the QSH regime each terminal $\alpha$ injects two modes that propagate towards opposite edges.

We consider the same setup as in Sec.~\ref{sec:IQH}, namely, that a $V_{\rm bias}$ is applied between the terminal $1$ and $6$ (see Fig.~\ref{fig:hall_bars_setup_A}) and that the remaining terminals act as voltages probes and, hence, $I_2=I_3=I_4=I_5=0$.
We use Eq.~(\ref{conductanceqsh}) to solve Eq.~(\ref{current}) for the unknown voltages to obtain
\begin{align}
\hspace{-0.04cm}
\left(
  V_1,  
  V_2,  
  V_3,  
  V_4,  
  V_5,  
  V_6
\right)
=
\left(
 \frac{3}{2}, 
 1,   
 \frac{1}{2}, 
 1,   
 \frac{1}{2}, 
0
\right) \frac{h}{e^2}I,
\label{voltagesqsh}
\end{align}
where $I=(2e^2/3h)V_{\rm bias}$, in agreement with previous papers \cite{Protogenov2013}.

The QSH transverse resistance is $R_{xy} =0$.  Since the voltage probes are not spin resolved, the transverse charge resistance vanishes due to the symmetry between the electronic propagation at the ``upper" and ``lower" edges of the Hall bar. 
Thus, there is no charge imbalance between the transverse terminals.
In turn, the longitudinal resistance is  $R_{xx}=h/2e^2$ and the contact resistance is $R_c=R_{16,16}-R_{xx}=h/e^2$. As expected for two resistors in series, $R_c$ is twice the resistance corresponding to a terminal with two perfectly conducting propagating modes.

It has been suggested that dephasing processes can explain why it has been so difficult to observe a perfect conductance quantization in QSH 
systems (see, for instance, the supplemental material of Ref. \cite{Roth2009} and Refs. \cite{Protogenov2013, Mani2016}).
Let us discuss this result, that has been mostly overlooked, using the B\"uttiker voltage probe model (see, for instance, Sec.~\ref{sec:QH_numerics}).
In IQH systems, chirality prevents a voltage probe to inject electrons back to the same edge state from which they have been drained, preventing interference effects between injected and drained electrons. 
In distinction, in the QSH regime a voltage probe that relaxes both momentum and spin and, thus, injects electrons at both the clockwise and anti-clockwise helical propagating edge states, promoting interference between counter propagating helical edge modes with opposite spins orientations.
The most relevant dephasing microscopic mechanisms for topological insulators are reviewed in Ref.~\cite{Qi2019}.

Using a three-terminal system \cite{Buttiker2009} with a strongly coupled voltage probe, $\mathcal G_{\alpha\varphi}=2e^2/h$, we find that the QSH effective conductance, Eq.~(\ref{conductanceeff}), drops from $2e^2/h$ to $3e^2/2h$.
Here the voltage probe interferes with one of the edge states carrying half of the current between the original two terminals, 
leading to an effective resistance of $2h/3e^2$ \cite{Protogenov2013,Mani2016}.
The supplemental material of Ref.~\cite{Roth2009}  studies weakly coupled voltage probes $\mathcal G_{\alpha\varphi} \ll e^2/h$ and models a gradual suppression of the perfect quantized conductance, a hallmark of the QSH, with increasing dephasing.

Let us now qualitatively discuss the nonequilibrium currents.
We take $E_F$ within the topological gap, a situation where only helical edge states are responsible for the electronic transport.

\begin{figure}[tbp]
\centering
\includegraphics[width=0.75\columnwidth]{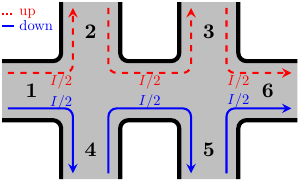} 
\caption{Local nonequilibrium electronic current map estimated using Eq.~\eqref{iklneq}. Here, the current $I=(2e^2/3h)V_{\rm bias}$ is established as a response of the bias $V_{\rm bias}$ for $E_F$ inside the topological gap.
The net spin of the electronic flow is indicated as dashed red for spin up and as solid blue for spin down.}
\label{fig:hall_bar_QSH_local_i}
\end{figure}

Using Eq.~\eqref{iklneq}, we build the nonequilibrium local current map shown in Fig.~\ref{fig:hall_bar_QSH_local_i} by weighting each local conductance map by its respective voltage, see Eq.~(\ref{voltagesqsh}). 
At the edge between terminals $1$ and $2$, only the local conductances from those terminals are not zero, with values $e^2/h$ per channel. 
Since $V_1=(3/2)(hI/e^2)$ and $V_2=(hI/e^2)$ Eq.~(\ref{iklneq}) gives a net local current of $I/2$ propagating from terminal $1$ to terminal $2$ with spin-up orientation. 
Repeating this reasoning over the whole system, we find that (i) all edges have current $I/2$, (ii) terminals $1$ and $6$ are indeed injecting and receiving a total current $I$ and (iii) terminals $2$ through $5$ are voltage probes with zero net current.  
Figure~\ref{fig:hall_bar_QSH_local_i} shows that there is no net spin current flowing from terminals $1$ to $6$, as expected in the absence of a spin polarized bias.

Let us now consider equilibrium currents.
Our discussion is based on  the local transmissions maps defined in Eq.~(\ref{localtcc}) for energies $E$ within the topological gap.
We postpone the quantitative analysis of the local equilibrium currents, that involve an integration over all occupied states, Eq.~(\ref{ikleq}), to the next subsection. 

\begin{figure}[tbp]
	\centering
		\includegraphics[width=0.49\columnwidth]{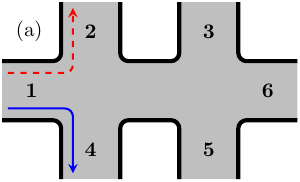} 
		\includegraphics[width=0.49\columnwidth]{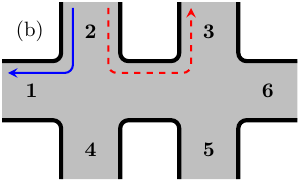} 
		\includegraphics[width=0.49\columnwidth]{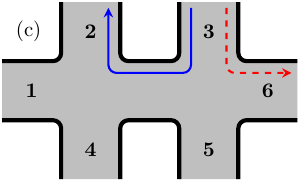} 
		\includegraphics[width=0.49\columnwidth]{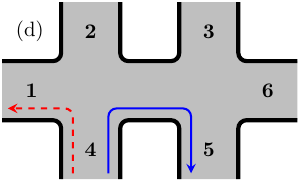} 
		\includegraphics[width=0.49\columnwidth]{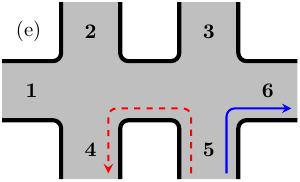} 
		\includegraphics[width=0.49\columnwidth]{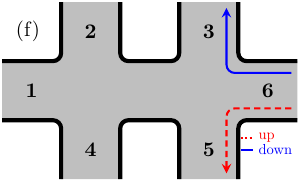} 
	\caption{ %
	Sketches of local transmissions $\widetilde T_{i\sigma,j\sigma}(E)$ for a Hall bar system in the QSH regime. 
	Panels (a) to (f) correspond to an electronic injection from the terminal $\alpha=1$ to $6$, respectively. 
	The spin-orbit coupling breaks the spin up (dashed red) and spin down (solid blue) degeneracy.
	}
		\label{fig:hall_bar_QSH_local_t}
\end{figure}

Figure~\ref{fig:hall_bar_QSH_local_t} shows the local charge and spin transmission $\widetilde T_{i\sigma,j\sigma}(E)$.
Each terminal injects two modes that have opposite spin orientations and propagate at opposite system edges towards the closest neighboring terminal.
Notice that, due to spin-momentum locking, the edge between terminals $1$ and $2$ supports an electronic spin-up propagation flowing from $\alpha=1$ to $2$ [Fig.~\ref{fig:hall_bar_QSH_local_t}(a)] and spin-down propagation from $2$ to $1$ [Fig.~\ref{fig:hall_bar_QSH_local_t}(b)]. 
Thus, the net charge transmission vanishes. 
The superposition of the transmission maps results in a zero net charge transmission, since  at all edges of the Hall bar there are counter propagating electronic states that cancel each other.
Therefore, the equilibrium contribution to the local charge current in Eq.~(\ref{ikleq}) vanishes.
In turn, our analysis suggests the presence of finite equilibrium edge spin currents circulating in the system. 

Our qualitative analysis also shows that local transmissions maps obtained considering single-terminal injections can be deceiving.
Only by taking into account all terminal injections one obtains the correct current.
This indicates that the pictures usually drawn in experimental studies of spin and valley Hall effects 
\cite{Mihajlovic2009,Balakrishnan2014,Avsar2014,Shimazaki2015,Katosh2015,Park2017,Safeer2019,Herling2020} 
can be misleading since they indicate that the current injected from the source terminal flow to neighboring terminals 
(depending on the spin or the valley degrees of freedom) other than the drain terminal. 
This would imply the violation of charge conservation.
The bias defines the source and drain terminals, the electronic current enters the system from the source terminal, travels through a path 
determined by interaction of the electrons with the material, and leaves the system at the drain terminal. 
In distinction, in linear response the local transmissions $\widetilde {\mathcal T}^{\beta}_{kl}$ do not contain information about the bias and the
electron probability to propagate toward different terminals depends on their spin, valley, and/or orbital quantum numbers.
$\widetilde {\mathcal T}^{\beta}_{kl}$
corresponds to 
the local current path only for the case of current injected from a single terminal $\beta$ ($V_\beta \neq 0$) with all the 
other terminals $\alpha$ ($\alpha\neq\beta$) grounded ($V_\alpha = 0$), see Eq.~(\ref{iklneq}).
For instance, in Fig.~\ref{fig:hall_bar_QSH_local_t}(a) $\widetilde {\mathcal T}^{\beta=1}_{kl}$ indicates that the current injected from terminal $1$ can propagate to both terminals $2$ and $4$, depending on the spin projection, and are not allowed to propagate to terminal $6$. Conversely, the actual local current $\widetilde{I}_{kl}^{}$ set by the bias voltage flows from terminal $1$ (source) to terminal $6$ (drain), as Fig.~\ref{fig:hall_bar_QSH_local_i} shows.
In summary, one must not confuse the actual electronic current path $\widetilde{I}_{kl}^{}$ with transmission probabilities paths $\widetilde {\mathcal T}^{\beta}_{kl}$.
The first is determined by the bias voltage, while the latter are not.

\subsection{Numerical results}
\label{sec:QSH_numerics}

For the numerical analysis of the local currents in the QSH regime we use the Kane-Mele (KM) model  \cite{Kane2005}, that describes the low-energy properties of electronic states in graphene with strong spin-orbit interaction by a tight-binding Hamiltonian, namely
\begin{align}
	H = -\sum_{\left<i,j\right>,\sigma} t_{1} c_{i\sigma}^\dagger c_{j\sigma} + \sum_{\left<\left<i,j\right>\right>,\sigma,\sigma'} it_2\nu_{ij}S^z_{\sigma\sigma'} c_{i\sigma}^\dagger c_{i\sigma'}.
	\label{hamiltonianresults}
\end{align}
The first term accounts for nearest-neighbors hopping processes with $t_{1}=t=2.7$ eV \cite{CastroNeto2009}.
The second term includes the spin-dependent hopping amplitude between second neighbors with $t_2=0.065t$.
The presence of spin-orbit coupling creates a topological gap $\Delta_{\rm G}=6\sqrt{3}t_2$ \cite{Kane2005}
such that the bulk bands onsets occur at the energies $\pm\Delta_{\rm G}/2=\pm 0.338t$.
Here $\mathbf S^z$ is the Pauli matrix $z$ component and $\nu_{ij}=+1$ ($\nu_{ij}=-1$) if the path from $i$ to $j$ follows the counterclockwise (clockwise) direction with respect to the hexagon centers of the honeycomb lattice. 
The graphene Hall bar is attached to six semi-infinite leads, where the spin-orbit coupling is turned off.
The leads are doped at $E=-t$ to maximize the density of propagating modes.

In order to compute both nonequilibrium and equilibrium currents, we employ the multiprobe recursive Green's function 
method \cite{Lima2018} to calculate the local spin resolved transmission coefficients given by Eq.~(\ref{localtcc}).
For simplicity, the leads are modeled by highly doped pristine graphene semi-infinite ribbons to maximize the electronic density of states and mimic metallic contacts used in standard experiments. A smoother and more realistic system-contact model is expected to blur the currents at the interface without qualitative changes on our results.

\begin{figure}[htbp]
	\centering
	\includegraphics[width=0.99\columnwidth]{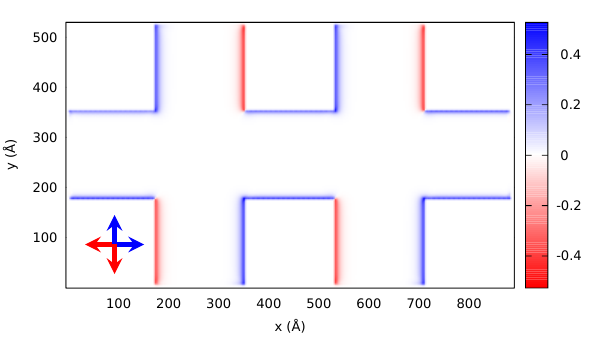} 
\caption{
{\it Nonequilibrium} local electronic charge current ${\bf \widetilde{I}}^{\rm neq}$ for the KM model Hamiltonian in the QSH regime 
for $E_F=-0.3t$.
The electronic current is presented in units of $e^2V_{\rm bias}/h$, includes both spin orientations, and is computed using Eq.~(\ref{iklneq}).
The color code is the same as in {Fig.~\ref{fig:IQH_eq}}.
}
	\label{fig:QSH_neq}
\end{figure}

Figure~\ref{fig:QSH_neq} shows the nonequilibrium component of the local current  $\widetilde{I}_{ij}^{\rm eq}$, obtained by computing Eq.~(\ref{iklneq}) for $E_F=-0.300t$. 
We plot the local charge current driven by $V_{\rm bias}$ (including both spin orientations).
As expected from the discussion in Sec.~\ref{sec:QSH}, see Fig.~\ref{fig:hall_bar_QSH_local_i}, the electronic transport from source (left) to drain (right) occurs through both top and bottom edge states.
For the chosen model parameters, the top (bottom) edge carriers has $\sigma=1$ ($\sigma=-1$).

Let us now address a more specific issue: The nonequilibrium currents at the system-terminal interfaces at the QSH and IQH regimes.
In the QSH regime, Fig.~\ref{fig:QSH_neq}, the current flowing through the central region crosses a system-terminal interface via one of the edges and it is injected back into the system at an opposite edge.
In this case, the interface has low resistance, so that all the electronic current crosses the interface at the position it arrives.
The strong current directionality fades away as the electrons enter the leads and populate trivial (non-QSH) states allowing
for bulk propagation connecting opposite edges. 
That is not the case in the IQH regime: Figure~\ref{fig:IQH_neq} clearly shows a finite current at the system-terminal interface connecting opposite edges.
In this case, we consider a partition where the magnetic field is absent in the leads and the interface has high resistance, so that only a fraction of the electronic current is absorbed at the position where it arrives and the remaining fraction continues to travel in the system.
As a result, the change in its direction happens partially inside the system itself and inside the leads as well.
In an experiment, the current is mostly likely to turn back to the system inside the voltage probe lead where the material constituting the leads usually behaves as a trivial metal, not sharing the same properties of the studied system.
The contrast between the QSH and IQH system-terminal interface currents ${\bf I}^{\rm neq}$  is particularly large for voltage probe terminals.
This behavior is allowed since there is no conservation rule requiring the current to be entirely absorbed by any voltage lead.
It is only required that the current collected by a voltage lead must be injected back.

The computation of the equilibrium currents is more involved, since it requires an energy integration of $\widetilde T_{i\sigma,j\sigma}^\alpha$ over all occupied states. 
As previously discussed, equilibrium charge currents are absent in systems with time-reversal symmetry.
This is explicitly manifest by the counter propagating helical edge states, a hallmark of the QSH regime.
We verify that this is indeed the case.
We run extensive computations of  $\widetilde T_{i\sigma,j\sigma}^\alpha(E)$ and find that  $\widetilde I_{i,j}^{\rm eq}$ is zero within numerical precision along all tested cross sections and energy values.
On the other hand, a finite spin-orbit interaction favors the appearance of equilibrium spin currents \cite{Souma2005}, that is $\widetilde T_{i\uparrow,j\uparrow}^\alpha(E)\neq \widetilde T_{i\downarrow,j\downarrow}^\alpha(E)$.
In what follows we discuss the properties of the latter in the QSH regime.

\begin{figure}[htbp]
	\centering
		\includegraphics[width=0.99\columnwidth]{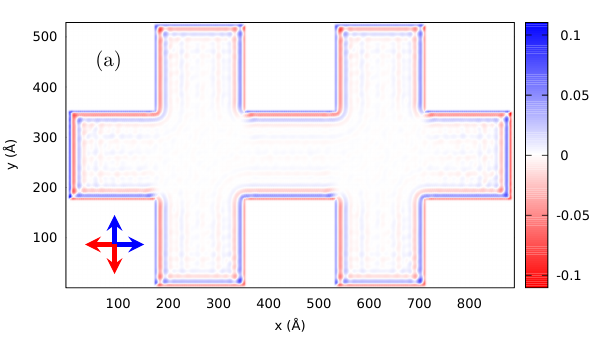}
		\includegraphics[width=0.99\columnwidth]{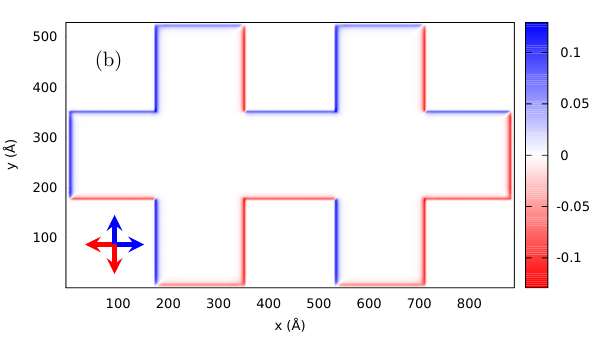}
	\caption{
	 {\it Equilibrium}  local  spin current $\Delta {\bf \widetilde{I}}^{s,{\rm eq}}$ in units of $t$ for the KM model Hamiltonian in the QSH regime calculated using 
	Eq.~(\ref{deltaikleqspin})  for (a) $\mu_1=-0.5t$, $\mu_2=-0.4t$ and (b) $\mu_1=-0.3t$, $\mu_2=-0.2t$.
	The color code is the same as in {Fig.~\ref{fig:IQH_eq}}.
	}
	\label{fig:QSH_eq}
\end{figure}

The equilibrium spin current, Eq.~\eqref{eq:I_spin_local}, is defined as the difference between the spin up and the spin down components of the equilibrium local charge current, namely,  
\begin{align}
\widetilde{I}_{ij}^{s,{\rm eq}} = \frac{\hbar}{2e}\left[\widetilde{I}_{i\uparrow,j\uparrow}^{\rm eq}-\widetilde{I}_{i\downarrow,j\downarrow}^{\rm eq}\right].
\end{align}  
As in the case of equilibrium charge current, an accurate evaluation of $\widetilde{I}_{ij}^{s,{\rm eq}}$ is difficult since it involves an integration over all occupied states. In analogy Eq.~(\ref{deltaikleq}), we define 
\begin{align}
	\Delta\widetilde{I}_{ij}^{\rm s,eq}(\mu_1,\mu_2) \equiv  \widetilde{I}_{ij}^{\rm s,eq}(\mu_2) - \widetilde{I}_{ij}^{\rm s,eq}(\mu_1).
	\label{deltaikleqspin}
\end{align}
According to the model parameters, the topological gap corresponds to $|E|< \Delta_{\rm G}/2 = 0.338t$.
That is, for $|E|>0.338t$ the electronic propagation is dominated by bulk states, while for $|E|<0.338t$ it occurs via edge states.
To capture information about the contributions from both edge and bulk propagation to the local equilibrium current,
we calculate $\Delta\widetilde{I}_{ij}^{\rm s,eq}(-0.5t,-0.4t)$ and $\Delta\widetilde{I}_{ij}^{\rm s,eq}(-0.3t,-0.2t)$.
We show the results in Fig.~\ref{fig:QSH_eq}. 

We find that the equilibrium spin current $\Delta\widetilde{I}_{ij}^{\rm s,eq}$ due to trivial states, see Fig.~\ref{fig:QSH_eq}(a), spreads over the Hall bar transverse cross sections and it is strongly enhanced at the system edges. Interference effects are also present, as discussed in Sec.~\ref{sec:QH_numerics}, when addressing local currents in the IQH regime, see Fig.~\ref{fig:IQH_neq}.
$\Delta\widetilde{I}_{ij}^{\rm s,eq}$ has a counter propagating structure along the horizontal direction.
This pattern originates from the superposition of bulk contributions to the local transmission taken at different energies. 
As a result, the equilibrium spin current shown in Fig.~\ref{fig:QSH_eq}(a) that flows clockwise at the system edges but alternates between clockwise and counter-clockwise propagation as the distance from the edges increases. 
Interestingly, even in the absence of helical edge states, the transport becomes stronger at the edges.
The equilibrium spin current in Fig.~\ref{fig:QSH_eq}(b) has only contributions from helical edge states. 
Here, the interference pattern dependence with the energy is very weak, since the dispersion of the helical edge states is linear, that is, the group velocity does not vary inside the topological gap.
As a result, the spin flow is narrower and stronger than the one in Fig.~\ref{fig:QSH_eq}(a) at the edges.
The circulation is clockwise and there is no current signal in the bulk. 

It is important to mention that previous papers \cite{Zheng2011,Chen2020} studying equilibrium currents in topological insulators have shown that  $\widetilde{I}_{ij}^{s,{\rm eq}}$ is zero at $E_F=0$.  This remarkable result has been obtained both analytically \cite{Zheng2011} and numerically \cite{Chen2020}  for isolated systems modeled by standard topological Hamiltonians. The analytical demonstration of this property relies on particle-hole symmetry \cite{Zheng2011}.
The QSH results presented here consider doped leads to simulate metallic contacts that break particle-hole symmetry.
Figures ~\ref{fig:QSH_eq}(a) and (b) do not allow to conclude that the local spin currents $\Delta\widetilde{{\bf I}}^{\rm s,eq}$ corresponding to trivial and topological states show a tendency to cancel each other.
We have tried to confirm if $\widetilde{I}_{ij}^{s,{\rm eq}}=0$ for $E_F=0$ using our formalism for the KM model with undoped leads and a two-terminal geometry, a much simpler setting than the Hall bar of Fig.~\ref{fig:hall_bars_setup_A}.
Unfortunately, the energy integral is still very difficult to converge to a precision that rules out a finite local current and the results we obtain are inconclusive. 
In real systems this issue looses importance, since particle-whole symmetry is absent in realistic materials electronic band structures and destroyed by disorder.

We conclude by stressing that, like in the IQH regime, the local equilibrium spin currents $\widetilde{I}^{s,{\rm eq}}_{ii'}$ cannot be detected by nonequilibrium transport experiments. Yet, one can envisage strategies to indirectly measure equilibrium currents, for instance, $\widetilde{I}^{s,{\rm eq}}_{ii'}$ carries a magnetic moment that gives origin to an electric dipolar field \cite{Chen2020}, which can in principle be measured. 

\section{Summary and conclusions}
\label{sec:conclusions}

We review the Landauer-B\"uttiker formalism to multiterminal systems and derive general expressions using nonequilibrium Green's functions that allows one to compute local electronic currents in mesoscopic systems at both the equilibrium and nonequilibrium regimes.
In distinction to the nonequilibrium transport that is dictated by the properties of the Fermi surface, equilibrium currents are governed by the system ground-state. 
We show that the results we put forward for the local electron currents can be projected into a suitable basis allowing the calculation of the spin-bond current. By analogy, this approach can be extended to obtain further transport properties, such as equilibrium and nonequilibrium valley and orbital local currents. 

To illustrate the formalism, we put forward a lattice toy model that consists of a central region with bonds forming a triangular ring subjected to a magnetic field that breaks the time-reversal symmetry. The simplicity of the model allows us to obtain closed analytical expressions for both equilibrium and nonequilibrium currents in an atomistic basis. These results clearly show that, in the absence of a bias voltage, the net electronic current through the system vanishes, but the net local (equilibrium) current is not necessarily zero, as previously found, for instance, in Ref.~\cite{Cresti2003} using NEGF.

Motivated by recent experiments \cite{Uri2020}, we further discuss the properties of equilibrium and nonequilibrium (linear response) local currents in the {IQH and the QSH} regime for a realistic Hall bar setup. 
As a case to point, we consider graphene systems. 

In the IQH regime, a nonvanishing local charge current is driven by the magnetic field induced broken time reversal symmetry.
We demonstrate that net equilibrium local charge currents circulate without leaving the system.
Our calculations show that any transverse section of the system has net equilibrium charge current equal to zero, as expected.
We revisit the voltage probe decoherence model to justify the absence of momentum relaxation that leads to the absence of longitudinal resistance.
We discuss the challenges involved in an accurate assessment of the energy integration in Eq.~\eqref{ikleq} for realistic model systems. To circumvent the latter, we propose a two-measurement protocol that avoids the necessity of accounting for all occupied states, and makes possible a quantitative comparison between theory and experiment.
 
In the QSH regime, the counter-propagating helical edge states interact at the system-terminal interfaces, forcing momentum relaxation and the appearance of a finite longitudinal resistance.
We show that at equilibrium, the local charge current vanishes due to time reversal symmetry, but the spin-orbit interaction gives rise to a finite local spin current that circulates the system with no net current flowing through any cross section of the system.
We find that our two-measurement protocol reveals the nature of trivial and topological equilibrium currents in QSH systems.

We emphasize the distinction between transmission, conductance and current.
Thus, we learn that local transmissions maps obtained
considering single terminal injections can be deceiving 
\cite{Mihajlovic2009,Balakrishnan2014,Avsar2014,Shimazaki2015,Katosh2015,Park2017,Safeer2019,Herling2020}.
Only by taking into account all terminal transmissions properly weighted by the terminal voltages one obtains meaningful results.

Finally, our results are nicely interpreted by the B\"uttiker picture \cite{Datta1995} that considers the injection of electrons by all terminals for electron energies ranging over the entire band, weighted by the Fermi distribution. In equilibrium, all terminals attached to the system simultaneously inject the same current, leading to a zero {\it net} equilibrium electronic current.
The presence of an external magnetic field breaks time-reversal symmetry and gives rise to nonvanishing dissipationless equilibrium {\it local} charge currents, while spin-orbit interactions {originate} nonvanishing dissipationless equilibrium local spin currents with zero local charge current.
Let us stress that despite not being detectable by nonequilibrium transport measurements, equilibrium currents can be assessed indirectly, as discussed in Secs. \ref{sec:IQH} and \ref{sec:QSH}.

We believe that the formalism we put forward and the results we obtain can be very helpful in  the understanding of future experiments on quantum imaging of current flow in different two-dimensional setups and material platforms.

\acknowledgments

We thank B. Nikoli\v{c} for enlightening discussions at the early stage of this work. We acknowledge financial support of the Brazilian Institute of Science and Technology (INCT) in Carbon Nanomaterials and the Brazilian agencies CAPES, CNPq, FAPEMIG, and FAPERJ.
The simulations were partially performed at the High Performance Computing Center (NACAD) at COPPE/Federal University
of Rio de Janeiro, Brazil.

\appendix

\section{Landauer-B\"uttiker equations}
\label{appendix:lbequations}

In this appendix we present alternative representations of the Landauer-B\"uttiker equation for multiterminal systems, discuss the value of the diagonal terms of the conductance matrix and the sign convention of the electronic current.

In Ref.~\cite{Buttiker1986}, B\"uttiker has assumed that each reservoir injects a positive current into the system through a terminal attached to it. 
In line with Ref.~\cite{Buttiker1986}, we consider that the positive current injected by the reservoir $\alpha$ into the system is given by 
$(e/h)\int dE\ \mathcal N_{\alpha}(E) f_\alpha(E)$ is accompanied by a negative current $-(e/h)\int dE\ \mathcal R_{\alpha}(E) f_\alpha(E)$
reflected back to reservoir $\alpha$.
Here $\mathcal N_{\alpha}(E)$ and $\mathcal R_{\alpha}(E)$ are the number of modes and the reflection coefficient at energy $E$ in lead $\alpha$, respectively \cite{Ihn2010}. 
The current injected from a given terminal $\beta$, with $\beta \neq \alpha$, reduces the current at  $\alpha$ by  
$-(e/h)\int dE\ \mathcal T_{\alpha\beta}(E) f_\beta(E)$,
where $\mathcal T_{\alpha\beta}$ is the transmission coefficient from $\beta$ to $\alpha$.
Thus, the net current at the terminal $\alpha$
is given by
\begin{align}
	I_{\alpha} =  \frac{e}{h}\int dE\ \bigg\{   
	& \left[\mathcal N_{\alpha}(E)  - \mathcal R_{\alpha}(E)\right] f_\alpha(E) 
\nonumber\\
	 & - \sum_{\beta\neq\alpha} \mathcal T_{\alpha\beta}(E) f_\beta(E)
	\bigg\}.
	\label{ialpha0}
\end{align}

In the thermodynamic equilibrium, where $f_\alpha(E)=f_\beta(E)$, the nonequilibrium current must be zero. 
Thus, the reflection and transmission probabilities must add to the number of modes, that is, $ \mathcal R_{\alpha}(E) + \sum_{\beta\neq\alpha} \mathcal T_{\alpha\beta}(E) = \mathcal N_{\alpha}(E)$. Hence, Eq.~(\ref{ialpha0}) can be cast as
\begin{align}
	I_{\alpha} = \sum_{\beta\neq\alpha} \frac{e}{h}\int dE\ 
	 \mathcal T_{\alpha\beta}(E) \left[f_\alpha(E) - f_\beta(E) \right].
	\label{ialpha1}
\end{align}
Terminal $\alpha$ injects positive current $I_\alpha$ into the system when $f_\alpha(E) > f_\beta(E)$, which occurs when the chemical potential $\mu_\alpha>\mu_\beta$. 

We use the standard chemical potential parametrization $\mu_\alpha = \mu_0+eV_\alpha$ \cite{Buttiker1986,Blanter2000,Hernandez2013},
$f_\alpha(E)=f_0(E)+(-\partial f_0/\partial E)eV_\alpha$ and Eq.~(\ref{ialpha1}) to obtain
\begin{align}
	I_{\alpha} = -\sum_{\beta\neq\alpha} \mathcal G_{\alpha\beta} \left(V_\beta - V_\alpha \right),
	\label{ialpha2}
\end{align}
where the conductance $\mathcal G_{\alpha\beta}$ is given by Eq.~(\ref{conductance0}) in the main text. 
Notice that applying a voltage $V_\alpha-V_\beta>0$ one creates a chemical potential difference $\mu_\alpha-\mu_\beta>0$ that induces a positive current $I_\alpha$, which means that Eq.~(\ref{ialpha2}) ensures a positive current flowing from the terminal with higher voltage to the terminal with lower voltage.

Defining a diagonal transmission probability $\mathcal T_{\alpha\alpha}(E) \equiv \mathcal R_{\alpha}(E) - \mathcal N_{\alpha}(E)$, Eq.~(\ref{ialpha0}) can also be cast as
\begin{align}
	I_{\alpha} = -\sum_{\beta} \frac{e}{h}\int dE\ \mathcal T_{\alpha\beta}(E) f_\beta(E),
	\label{ialpha3}
\end{align}
where the sum runs through all the terminals, including $\beta=\alpha$.
The expansion of $f_\beta$ in Eq.~(\ref{ialpha3}) up to linear order in the voltage $V_\beta$ gives
\begin{align}
	I_{\alpha} = 
	- \frac{e}{h}\int \!dE \! \left[\sum_{\beta} \mathcal T_{\alpha\beta}(E)\right] \! f_0(E) 
	-\sum_{\beta} \mathcal G_{\alpha\beta} V_\beta.
	\label{ialpha4}
\end{align}

Since charge conservation requires that the total current in the system must sum up to zero and there is
no current flow if all the voltages are the same, one finds the sum rules \cite{Buttiker1986,Ihn2010}
\begin{align}
	\sum_{\beta} \mathcal T_{\alpha\beta}(E) =
	\sum_\alpha \mathcal G_{\alpha\beta} =\sum_\beta \mathcal G_{\alpha\beta} =0,
	\label{lbconservationrules}
\end{align}
which allows us to write Eq.~(\ref{ialpha4}) as
\begin{align}
	I_{\alpha} = -\sum_{\beta} \mathcal G_{\alpha\beta} V_\beta.
	\label{ialpha5}
\end{align}

From the conservation rules in Eq.~(\ref{lbconservationrules}) we calculate the diagonal elements of the transmission and the conductance matrices as
\begin{align}
\mathcal T_{\alpha\alpha} 
	&= -\sum_{{\beta\ne\alpha}}\mathcal T_{\alpha\beta}, \\
	\mathcal G_{\alpha\alpha} 
	&= -\sum_{{\beta\ne\alpha}}\mathcal G_{\alpha\beta}
	 = -\sum_{{\alpha\ne\beta}}\mathcal G_{\alpha\beta},
\end{align}
respectively.

We can also rewrite Eq.~(\ref{ialpha5}) using Eq.~(\ref{lbconservationrules}) to recover Eq.~(\ref{ialpha2}) as
\begin{align}
I_\alpha 
&= - \sum_{\beta\neq\alpha} G_{\alpha\beta} V_\beta - \left(-\sum_{\beta\neq\alpha} G_{\alpha\beta}\right) V_\alpha  \\
&= -\sum_{\beta\neq\alpha} G_{\alpha\beta} \left(V_\beta - V_\alpha\right).
\end{align}
Therefore, the Landauer-B\"utikker equation can be cast by Eqs.~(\ref{ialpha5}) or (\ref{ialpha2}), since both are completely equivalent.

We stress that the negative sign in Eq.~(\ref{ialpha5}) implies a positive current flowing from the terminal with the highest voltage to the one with the lowest voltage. 
A positive voltage $V_\beta$ in Eq.~(\ref{ialpha5}) renders a negative contribution to the current $I_\alpha$ because the conductance $\mathcal G_{\alpha\beta}$ from $\beta$ to $\alpha$ is positive.
On the other hand, a positive voltage $V_\alpha$ renders a positive contribution to $I_\alpha$ because $\mathcal G_{\alpha\alpha}$ is a negative quantity.
For instance, assume that a bias voltage $V_{\rm bias}$ is applied between terminals $1$ ($V_1=V_{\rm bias}$) and $2$ ($V_2=0$) in a two-terminal system and that the conductance matrix elements read $\mathcal G_{12}=\mathcal G_{21}=e^2/h$ and  $\mathcal G_{11}=\mathcal G_{22}=-e^2/h$. 
Thus, Eq.~(\ref{ialpha5}) leads to 
$I_1=-\mathcal G_{11}V_1 - \mathcal G_{12}V_2 = (e^2/h) V_{\rm bias}>0$ and
$I_2=-\mathcal G_{21}V_1 - \mathcal G_{22}V_2 =-(e^2/h) V_{\rm bias}<0$.

Equation~(\ref{ialpha2}) is less often used, but has the advantage that the current is explicitly written in terms of voltage differences, which also guarantees that we can set one of the voltages to zero without loss of generality \cite{Datta1995}.
It is simpler and straightforward to obtain the current in the two-terminal system, namely, 
$I_1 = \mathcal G_{12} (V_1-V_2) = (e^2/h) V_{\rm bias}$ and 
$I_2 = \mathcal G_{21} (V_2-V_1) =-(e^2/h) V_{\rm bias}$.
Moreover, in multiterminal systems, Eq.~(\ref{ialpha2}) provides a simple way to calculate the voltage measured by a voltage probe, where the terminal has zero net current. 
Assuming that the voltage probe is at the terminal $\gamma$ ($I_\gamma=0$), Eq.~(\ref{ialpha2}) yields \cite{Datta1995,Ihn2010}
\begin{align}
	V_\gamma = \frac{\sum_{\beta\neq\gamma} G_{\gamma\beta} V_\beta }{\sum_{\beta\neq\gamma} G_{\gamma\beta}}.
\end{align}
The use of voltage probes is frequent in experiments and calculations aiming to obtain longitudinal and transverse resistances, as described in the main text.

We also emphasize that some authors take into account the negative charge of the carriers and define as negative the charge current being injected from the reservoir to the leads \cite{Ihn2010}.
As a consequence, $I_\alpha$ has the opposite sign of Eq.~(\ref{ialpha0}).
In this case, one must consider the action of a positive voltage on the chemical potential of negative charges as $\mu_\alpha=\mu_0-eV_\alpha$, which changes the
sign of the linear term on $V_\alpha$ of the Fermi-Dirac distribution approximation.
Thus, one obtains versions of Eqs.~(\ref{ialpha2}) and (\ref{ialpha5}) without the minus signs, namely, $I_\alpha = \sum_{\beta} \mathcal G_{\alpha\beta} (V_\beta-V_\alpha)$ and $I_\alpha = \sum_{\beta} \mathcal G_{\alpha\beta} V_\beta$, respectively, that satisfy the conservation rules in Eq.~(\ref{lbconservationrules}).
In this picture, a positive bias voltage $V_\beta-V_\alpha>0$ indicates a negative chemical potential difference $\mu_\beta-\mu_\alpha<0$.
Electrons will flow from the reservoir with higher chemical potential $\mu_\alpha$ to the one with lower chemical potential $\mu_\beta$, establishing a positive current from terminal $\beta$ (higher potential $V_\beta$) to terminal $\alpha$ (lower potential $V_\alpha$), in agreement with Ohm's law.

\section{Toy model Green's functions and transmissions in matrix representation}
\label{app:gfsandtransmissions}

In this appendix we present the explicit matrix representation of some standard useful formula that appear in the main text. 

The retarded Green's functions reads
\begin{align}
	\mathbf G^r = \left(E\mathbf 1-\mathbf H-\mathbf \Sigma^r\right)^{-1} = \frac{1}{D} \mathbf C^T,
\end{align}
where $D\equiv \text{det}\left(E\mathbf 1-\mathbf H-\mathbf \Sigma\right)$ and $\mathbf C$ is the co-factor 
matrix of $E\mathbf 1-\mathbf H-\mathbf \Sigma$ with elements $C_{ll'} = (-1)^{l+l'} M_{ll'}$. 
The minor $M_{ll'}$ is the determinant of $E\mathbf 1-\mathbf H-\mathbf \Sigma^r$ without the row $l$ and the column $l'$.
Analogously, the advanced Green's functions is $\mathbf G^a = \left(\mathbf G^r\right)^\dagger =\frac{1}{D^*} \mathbf C^*$.
Thus, 
\begin{align}
	\mathbf G^r = \frac{1}{D  } \mathbf C^T
	\quad \text{and} \quad
	\mathbf G^a = \frac{1}{D^*} \mathbf C^*.
\end{align}

In turn, the spectral function $\mathbf A_\alpha \equiv \mathbf G^r \mathbf \Gamma_\alpha	\mathbf G^a$ reads
\begin{align}
	A_{\alpha,lk} 
&	= \sum_j G^r_{lj} \Gamma_{\alpha,jj}	G^a_{jk}
	= \frac{1}{|D|^2} \sum_j C_{\nu'\nu} \Gamma_{\alpha,jj} C_{kl'}^* \nonumber\\
&	= \frac{1}{|D|^2} \sum_j (-1)^{l+j}M_{jl} \Gamma_{\alpha,jj} (-1)^{j+k}M_{jk}^* \nonumber\\
&	= \frac{(-1)^{l+k}}{|D|^2} \sum_k M_{jl} \Gamma_{\alpha,jj} M_{jk}^*.
\end{align}
Notice that $ A_{\alpha,ll} 	= \frac{1}{|D|^2} \sum_j \Gamma_{\alpha,jj} \left|M_{jl}\right|^2 \ge 0$.

The local transmission, defined by Eq.~(\ref{localtcc}), becomes
\begin{align}
	\widetilde{\cal T}_{\nu'\nu}^{\alpha}(E) \equiv
	2\frac{(-1)^{l+k}}{|D|^2} \sum_j \text{Im}\left[  M_{jl} \Gamma_{\alpha,jj} M_{jk}^*  H_{\nu'\nu} \right].
					\label{localtcc2}
\end{align}
We write the total transmission, Eq.~\eqref{eq:transmission}, as
\begin{align}
	{\cal T}_{\alpha\beta} 
&	
= \sum_{\nu'\nu} \Gamma_{\alpha,kk} G^r_{\nu'\nu} \Gamma_{\beta,ll}	G^a_{\nu\nu'} 
\nonumber\\ &	
= \frac{1}{|D|^2} \sum_{\nu'\nu} \Gamma_{\alpha,ll} C_{\nu\nu'} \Gamma_{\beta,kk}	C^*_{\nu\nu'}, 
\nonumber\\
&	= \frac{1}{|D|^2} \sum_{\nu'\nu} \Gamma_{\alpha,ll} \left|M_{\nu\nu'}\right|^2 \Gamma_{\beta,kk} \ge 0.
\label{totaltransmissionminor}
\end{align}

For simplicity, we use diagonal  line widths compatible with the toy model in Sec.~\ref{sec:toymodel} where the formulas are applied.
In order to obtain general expressions, one needs to account for the off-diagonal matrix elements of the line-widths.

\section{Toy model local transmission sign convention}
\label{app:toymodel-sign-convention}

Here we discuss the sign convention of the local transmission in Eq.~(\ref{localtcc}) using the toy model shown in Fig.~\ref{fig:toymodel}.
We consider that the site $2$ is disconnected from the other sites and that the magnetic field is zero ($t_{21}=t_{23}=\phi=0$).
In this case, the total transmission $\mathcal T_{RL}$ injected from left $L$ to right $R$ must equal the local transmission $\widetilde{\mathcal T}^{L}_{31}$ from site $1$ to site $3$ upon injecting from $L$.
Using Eq.~(\ref{localtcc2}) we find $\widetilde{\mathcal T}^{L}_{31}=2\,\text{Im}\left[M_{11}\Gamma_1M_{13}H_{31}\right]/\left|D\right|^2$, where $H_{31}=-t_{31}$ and the minors $M_{11}=E^2+iE\Gamma_3/2$ and $M_{13}=-Et_{13}$ are calculated using the matrix $E\mathbf 1 - \mathbf H - \mathbf \Sigma^r$ in Eq.~(\ref{ehstoymodel}).
Thus, $\widetilde{\mathcal T}^{L}_{31}=\Gamma_1\Gamma_3E^2\left|t_{13}\right|^2/\left|D\right|^2>0$.
In the case $t_{21}=t_{23}=\phi=0$, Eq.~(\ref{trltoymodel}) yields
$\mathcal T_{RL}=\Gamma_1\Gamma_3E^2t_{13}^2/\left|D\right|^2=\widetilde{\mathcal T}^{L}_{31}$.
Moreover, an analogous calculation shows that $\widetilde{\mathcal T}^{R}_{13}=\mathcal T_{LR}$.
Therefore, Eqs.~(\ref{eq:transmission}) and (\ref{localtcc}) follow the same sign convention.

Now we discuss how the local current responds to a positive chemical potential difference and a positive bias voltage.
We use Eq.~(\ref{localcc2}) to calculate the current from site $1$ to $3$, namely,
	$\widetilde{I}_{31}
	=  (e/h)\int\! dE [f_L(E) \widetilde{\mathcal T}_{31}^{L}(E) + f_R(E) \widetilde{\mathcal T}_{31}^{R}(E)]$.
Since $\widetilde{\mathcal T}^{L}_{31}>0$ and $\widetilde{\mathcal T}^{R}_{31}=-\widetilde{\mathcal T}^{L}_{31}<0$, 
the current is positive if $\mu_L>\mu_R$, negative if $\mu_L<\mu_R$ and vanish in the equilibrium $\mu_L=\mu_R$, as expected 
for systems with time-reversal symmetry.

In an analogous fashion, {we consider} a bias voltage $V_{\rm bias}>0$, where $V_L=V_{\rm bias}$ and $V_R=0$, 
{and use Eq.~(\ref{iklneq}) to calculate the nonequilibrium current to obtain $\widetilde{I}^{\rm neq}_{31}=\widetilde{\mathcal G}_{31}^{L} V_{\rm bias}$.} 
Since $\widetilde{\mathcal T}^{L}_{31}>0$, both the conductance $\widetilde{\mathcal G}_{31}^{L}$, given by Eq.~(\ref{conductancelocal}), and $\widetilde{I}^{\rm neq}_{31}$ are positive.
Therefore, the local current flows from site $1$ to site $3$, that is, it flows from the highest voltage (left) to the lowest voltage (right), as defined for the total current between terminals.

\section{Local equilibrium currents in graphene nanoribbons}
\label{app:acgnr}

\begin{figure}[tbp]
	\centering
		\includegraphics[width=0.85\columnwidth]{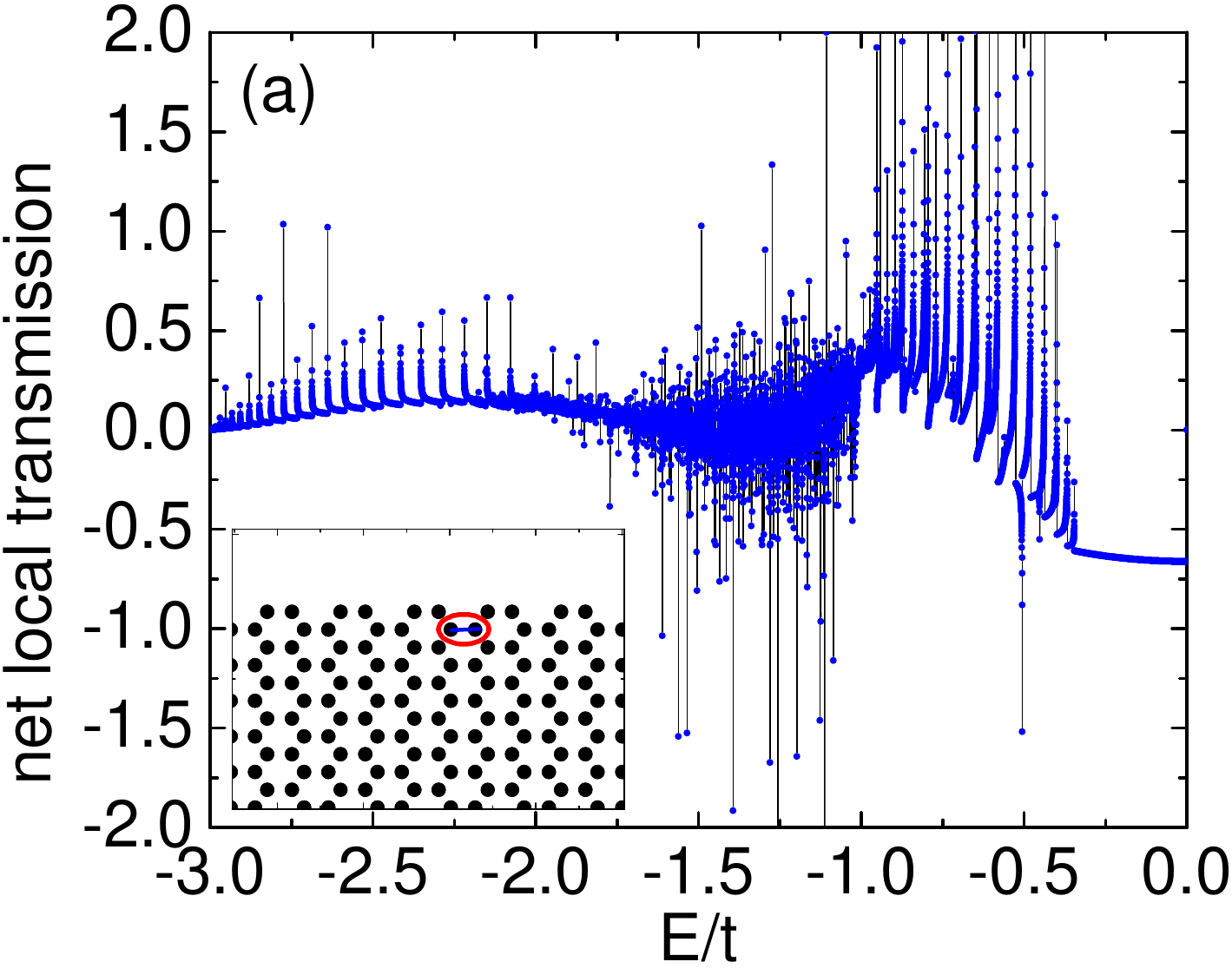}
		\includegraphics[width=0.85\columnwidth]{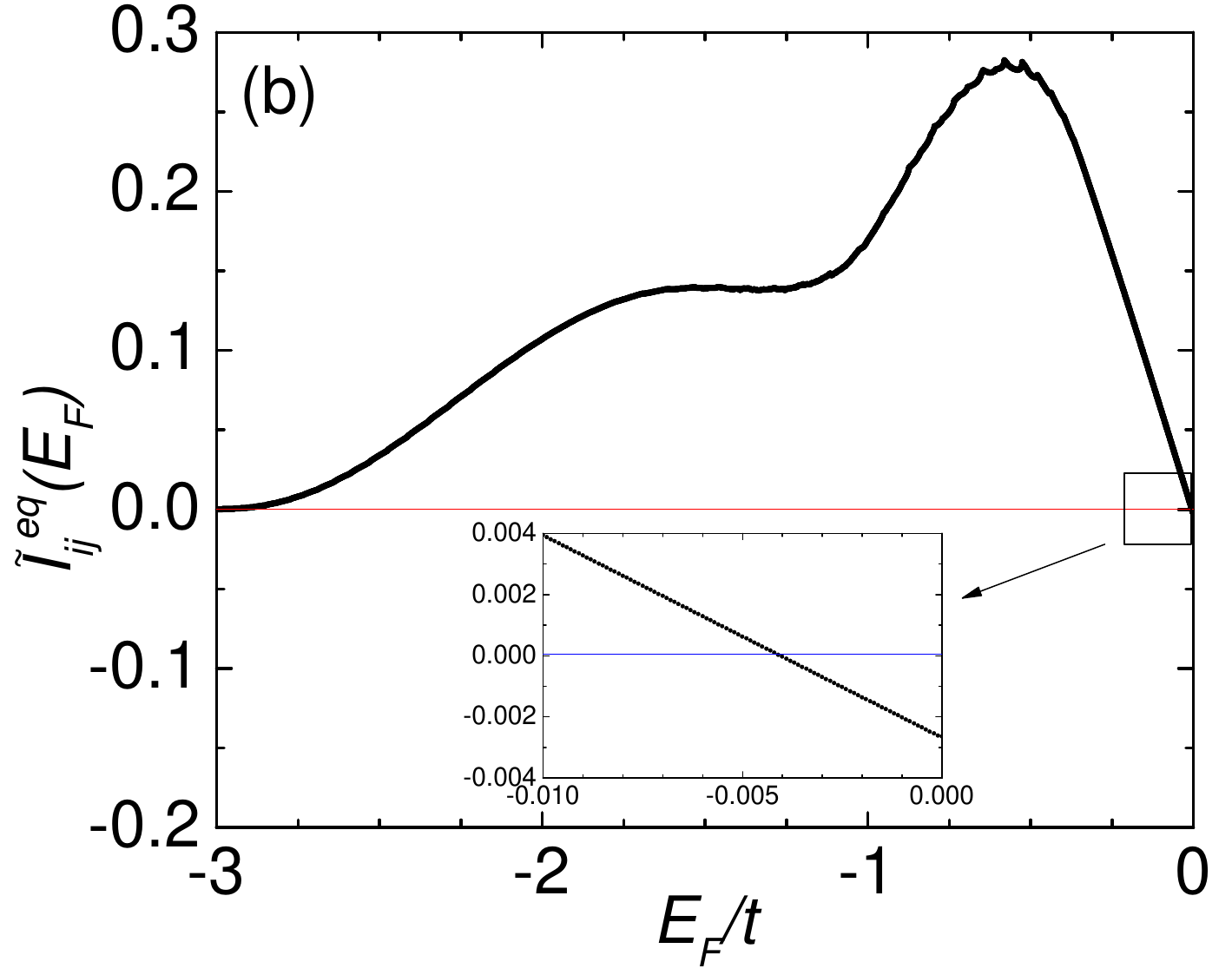}
	\caption{
	(a) Net local transmission as a function of the energy for a selected bond at the edge of a armchair graphene nanoribbon. The inset shows the selected bond $(i,j)$ by the red ellipse. (b) Local equilibrium current $\widetilde{{I}}_{ij}^{\rm eq}$ as a function of the Fermi energy $E_F$ ($T=0$) in units of $et/h$. The inset shows the deviation from zero for $E_F$ values approaching the charge neutrality point.
	}
	\label{fig:intribbon}
\end{figure}

As discussed in Sec.~\ref{sec:QH_numerics}, 
an accurate evaluation of the local equilibrium currents $\widetilde{{\bf I}}^{\rm eq}$, given by Eq.~(\ref{ikleq}), is a quite numerical challenge, 
that requires the integration of the net local transmission which is not a smooth function for all bonds over all system occupied states.
To illustrate the numerical difficulty, we present a quantitative analysis of the local equilibrium currents for a pristine armchair graphene nanoribbon 
with strong spin-orbit interaction in the QSH regime at $T=0$ using the model Hamiltonian defined in Sec. \ref{sec:QSH_numerics}.
For simplicity, we chose to analyze a two-terminal system, which renders a much faster computation then the multiprobe Hall bars, studied in Secs.~\ref{sec:IQH} and \ref{sec:QSH}.

We consider a $90$-\AA\ wide graphene nanoribbon, which has roughly the minimum width to prevent the overlap of opposite edge states, destroying the QSH regime.
Figure \ref{fig:intribbon}(a) shows the net local transmission $\sum_{\alpha=L,R}{\cal T}^\alpha_{ij}(E)$, the integrand of Eq.~(\ref{ikleq}), for a selected bond $(i,j)$ (indicated in red in the inset) near the edge as a function of the electronic energy $E$.
The integrand displays strong fluctuations in energy window corresponding to bulk (trivial) states. It presents sharp discontinuities at energies corresponding to the opening of transversal propagating modes, that are similar to Van Hove singularities found in the system density of states of quasi-1D systems and shows strong oscillations corresponding to interference effects that we attribute to influence of the leads. The net local transmission becomes a smooth function of $E$ within the topological gap located around the charge neutrality point.  

Figure~\ref{fig:intribbon}(b) gives  $\widetilde{{\bf I}}^{\rm eq}$ for the selected bond as a function of the Fermi energy $E_F$, it corresponds to the integral of  $\sum_{\alpha=L,R}{\cal T}^\alpha_{ij}(E)$ from the bottom of the band at roughly $-3t$ up to $E_F$. The integral behaves smoothly and tends to zero at the charge neutrality point $E_F=0$, which is in {\it qualitative} agreement with previous papers \cite{Zheng2011,Chen2020}. We note that by increasing the number of integration points $\widetilde{{I}}_{ij}^{\rm eq}(E_F=0)$ becomes rather small (as compared with its maximum value), but the convergence is rather slow. Due to the singularities of the integrand, see Fig.~\ref{fig:intribbon}(a), even using $10^4 \cdots 10^5$ integration points, we find $|\widetilde{{I}}_{ij}^{\rm eq}(E_F=0)| \alt 10^{-3} et/h$. 
The characteristic behavior of  $\widetilde{{I}}_{ij}^{\rm eq}(E_F)$ around $E_F=0$ is illustrated by the inset of Fig.~\ref{fig:intribbon}(b) where we have used $30000$ integration points. At this point it is important to recall that the typical accuracy of the recursive Green's functions method is of the order $10^{-7}$ for the conductance \cite{Lewenkopf2013,Lima2018}. 
Thus, the accurate computation of the local transmission in Eq.~(\ref{ikleq}) within numerical precision requires a larger number of points, which for larger systems is very time consuming and becomes rapidly prohibitive. 

This discussion gives further support for using the two-measurement protocol we propose, in order to compare theory with experiment and for considering chemical potential intervals over which the local transmissions show none or only few singularities.

\bibliography{equilibrium_currents}

\end{document}